\documentclass[12pt]{iopart}
\usepackage[utf8]{inputenc}
\usepackage{mdframed}
\usepackage{graphicx}
\usepackage{amssymb}
\usepackage{amsfonts,stmaryrd}
\usepackage{mathrsfs}
\usepackage{bm}
\usepackage[colorlinks=true]{hyperref}
\usepackage{iopams}
\usepackage{epsfig}
\usepackage{xcolor}

\newcommand{\CD}{{\cal D}}
\newcommand{\CQ}{{\cal Q}}

\def\U{\mathcal U}

\newcommand{\CR}{{\RicciScalR}}

\newcommand{\CC}{{\cal C}}
\newcommand{\s}{{\mathfrak S}}

\newcommand{\average}[1]{\left\langle #1 \right\rangle_\CD}

\newcommand{\dotaverage}[1]{\left\langle #1 \right\rangle^{\bdot}_\CD}
\newcommand{\ddotaverage}[1]{\left\langle #1 \right\rangle^{\bdot\bdot}_\CD}

\newcommand{\inI}{{\rm I}}
\newcommand{\inII}{{\rm II}}
\newcommand{\inIII}{{\rm III}}
\newcommand{\R}{\mathbb{R}}
\newcommand{\g}{\mathbf{g}}
\newcommand{\GBC}{\mathbf{G}}
\newcommand{\TT}{\mathbf{T}}

\newcommand{\initial}[1]{{#1_{\mathbf i}}}
\DeclareMathAlphabet\mathbfcal{OMS}{cmsy}{b}{n}

\def\D{\mathcal{D}}
\def\d{\mathrm{d}}

\def\TVolL{\boldsymbol\varpi}
\def\TVolR{\boldsymbol{\eta}}
\def\CVolL{\varpi}
\def\CVolR{\eta}
\def\RicciScalR{\mathcal{R}}
\def\M{\mathcal{M}}
\def\RicciScalL{\mathrm{R}}
\def\CRiemL{\mathrm{Riem}}
\def\TRiemL{\mathbf{Riem}}
\def\CRiemR{\mathcal{R}iem}
\def\TRiemR{\boldsymbol{\mathcal{R}iem}}
\def\CRicL{\mathrm{Ric}}
\def\TRicL{\mathbf{Ric}}
\def\CRicR{\mathcal{R}ic}
\def\TRicR{\boldsymbol{\mathcal{R}ic}}
\def\TWeyl{{\mathbf{C}}}
\def\CWeyl{\mathrm{C}}
\def\CmetricL{g}
\def\TmetricL{\mathbf{g}}
\def\CmetricR{h}
\def\TmetricR{\mathbf{h}}
\def\TSchouten{\mathbf{S}}
\def\CSchouten{S}

\def\CalA{\mathcal{A}}

\newtheorem{theo}{\bf{Theorem}}
\newtheorem{lem}{\bf{Lemma}}
\newtheorem{cor}{\bf{Corollary}}
\newtheorem{defi}{Definition}
\hypersetup{
    colorlinks,%
    citecolor=blue,%
    filecolor=blue,%
    linkcolor=magenta,%
    urlcolor=blue
}
\definecolor{MyGreen}{rgb}{0.0,.5,0.0}

\definecolor{MyDarkRed}{rgb}{0.7,0,0}

\definecolor{LeoRed}{rgb}{0.7,0,0.5}

\makeatletter
\long\def\@makefntext#1{\parindent 1em\noindent 
 \makebox[1em][l]{\footnotesize\rm$\m@th{^{\arabic{footnote}}}$}%
 \footnotesize\rm #1}
\def\@makefnmark{\hbox{$^{\arabic{footnote}}\m@th$}}
\def\@thefnmark{\arabic{footnote}}
\makeatother

\begin{document}
\setcounter{footnote}{0}
\title[Gauss--Bonnet--Chern approach to the averaged Universe]{Gauss--Bonnet--Chern approach to the \\averaged Universe}
\author{L\'eo Brunswic and Thomas Buchert}
\address{Univ Lyon, Ens de Lyon, Univ Lyon1, CNRS, Centre de Recherche Astrophysique de Lyon UMR5574, F--69007, Lyon, France
\\Emails:  leo.brunswic@ens-lyon.fr $\ and \ $ buchert@ens-lyon.fr}
\begin{abstract}
The standard model of cosmology with  postulated dark energy and dark matter sources may be considered as a fairly successful fitting model to observational data. However, this model leaves the question of the physical origin of these dark components open. Fully relativistic contributions that act like dark energy on large scales and like dark matter on smaller scales can be found through generalization of the standard model by spatially averaging the inhomogeneous Universe within general relativity.
The spatially averaged $3+1$ Einstein equations are effective balance equations that need a closure condition. Heading for closure we here explore topological constraints. Results are straightforwardly obtained for averaged $2+1$ model universes. For the relevant $3+1$ case, we employ a method based on the Gauss-Bonnet-Chern theorem generalized to Lorentzian spacetimes and implement a sandwich approach to obtain spatial average properties.
The $3+1$ topological approach supplies us with a new equation linking evolution of scalar invariants of the expansion tensor to the norm of the Weyl tensor. 
From this we derive general evolution equations for averaged scalar curvature and kinematical backreaction, and we discuss related evolution equations on this level of the hierarchy of averaged equations. We also discuss the relation between topological properties of cosmological manifolds and dynamical topology change, e.g. as resulting from the formation of black holes.
\end{abstract}

%
\section{Introduction}
\label{intro}
We first give the cosmological context, discuss the averaging of inhomogeneous cosmologies in general relativity,
outline the strategy toward the aim to close the system of averaged equations, and summarize the assumptions, aims and methods of this paper.

\subsection{The Dark Universe and the role of averaged model universes}

The standard model of cosmology assumes that the effective evolution of the Universe is governed by the Friedmann--Lema\^\i tre
equations for an idealized (homogeneous--isotropic) matter distribution, equipped with the Robertson--Walker metric of classical general relativity (henceforth FLRW model).
This idealization relies on the conjecture that the inhomogeneous Universe can be described, on average, by this class of solutions. A subcase is the concordance model assuming spatially flat space sections and a positive cosmological constant modeling dark energy. Although there is no physical justification for this model to provide the average behaviour of the Universe,
it matches a fair fraction of observational data\footnote{See, however, existing `tensions' with this model \cite{tensions}.}
if the energy density of the FLRW model is assumed to consist of
roughly $25$ percent of \textit{dark matter} and $70$ percent of \textit{dark energy}. The missing $5$ percent are attributed to known physical components like baryonic matter, neutrinos and radiation.

Relaxing this idealized description is challenging, since it involves improvement of the cosmological model by taking into account the effect of structure on the kinematic and geometric properties of model universes in the large. Strategies that face this challenge have to deal with the averaging problem that, in a simple realization, replaces the standard laws by their (spatially) averaged general counterparts. To realize these counterparts it is sufficient to (i) choose a foliation of spacetime, (ii) define an average operator and (iii) average the scalar parts of Einstein's equations
on the hypersurfaces of the foliation. A further simplification is then to choose an irrotational dust model and introduce Riemannian volume integration of scalar variables as a spatial average with respect to free-falling observers in a
flow-orthogonal slicing of spacetime. Relaxing these assumptions is possible, e.g. the resulting equations for a general fluid volume-averaged on slices of a general foliation that is tilted with respect to the fluid flow are not very different \cite{foliations,buchert:generalfluid,covariance}. We shall, as a first step and also because most of the literature in this context is based on the flow-orthogonal approach, employ the set of cosmological equations introduced in \cite{buchert:dust}, \cite{buchert:perfectfluid}.

As a natural consequence of such an averaging strategy, the cosmological equations feature additional fluctuating
parts---so-called \textit{backreaction terms}---condensed into a non-local
\textit{kinematical backreaction term} (composed of extrinsic curvature invariants) and a \textit{curvature deviation} (defined as the departure of the averaged intrinsic scalar curvature from the Friedmannian constant curvature model).  The relevant mechanism that qualitatively explains the emergence of kinematic dark matter and dark energy in this scheme is a coupling of the two fluctuating parts that is absent in the standard cosmological model. The `dark' sources are, in this scheme, a property of the geometry of spatial sections. They can be defined as  `intrinsic curvature energies', a positive intrinsic curvature mimicking \textit{dark matter} (expected for collapsing domains \cite{quentin:darkmatter,boudjan}), and a negative intrinsic curvature mimicking \textit{dark energy} (expected for a void-dominated Universe), and as a consequence of the non-conservation of averaged intrinsic curvature \cite{buchertcarfora:curvature}, this latter property is global. (For details the reader is directed to the reviews \cite{buchert:review,buchert:focus,ellisreview,buchertrasanen,whatisdust,buchert11}.)

\subsection{The closure issue and the methods employed in this paper}
\label{closureliterature}

Although the `dark' components can be explained qualitatively as effective properties of the inhomogeneous geometry, a quantitative assessment is needed, but hard to investigate. The ways explored have been mostly analytical and, recently, general-relativistic simulations have begun to be developed. The averaged inhomogeneous equations provide balance equations that need a \textit{closure condition}.
Analytical work has focused on the quantification of \textit{cosmological backreaction} by different assumptions on this closure condition. Firstly, modeling inhomogeneities by exact solutions of Einstein's equations (e.g., \cite[sect.7]{buchert:focus}, \cite{bolejko,sussman1,sussman2,zimdahl:LTB,sussman3}), or generic approximation methods based on the relativistic Lagrangian perturbation theory (\cite{RZA_1,RZA_2} and follow-up papers) provide such a closure.

Alternatively, one may focus on equations of state. This route follows the analogous procedure of standard cosmology: the equations of Friedmann and Lema\^\i tre for a perfect fluid source with pressure are not closed and an equation of state for the matter source is needed; similarly, the averaged cosmologies can be written in the form of these standard cosmological equations featuring \textit{effective} density and pressure sources that have to be related through a \textit{dynamical equation of state}. Examples are a Chaplygin equation of state
\cite{chaplygin}, or exact scaling solutions \cite{morphon,scaling}, a global stationarity assumption \cite{buchert:darkenergy,buchert:global}, or mappings of the backreaction fluid to other theories
e.g. to the modeling of bulk viscosity \cite{zimdahl:viscosity}, or assumptions on the form of the self-interaction potential in the scalar field analogy of a backreaction fluid, applied to scaling quintessence models \cite{morphon}, to scalar dark matter \cite{quentin:darkmatter}, and to inflation \cite{inflation}.

In this paper, we investigate a new theoretical path heading for closure by topological constraints,
based on the Gauss-Bonnet-Chern (GBC) theorem \cite{Chern1,Bredon,Kobayashi-Nomizu1} and on topological assumptions on the spacetime foliation.
The Gauss-Bonnet-Chern theorem relates the Euler characteristic, which is purely topological, to the average of intrinsic curvature invariants. 
The $2+1$ case can be completely determined following Magni's {Master} thesis work \cite{Magni} in which he obtains a  topological constraint by applying the Gauss-Bonnet theorem
to a space slice which is two-dimensional.
This case surely forms a toy model and one should be cautious {when} generalizing $2+1$ global backreaction to the $3+1$ setting. Indeed, the GBC theorem states that the total intrinsic curvature is conserved {only} in the former case while, in the latter case, the Gauss-Bonnet-Chern theorem applied to a three-dimensional spacelike hypersurface 
does not link geometrical quantities to topological invariants.
This difficulty is avoided by considering a tube of spacetime, not simply a leaf of a spatial foliation, and by performing a sandwich limit (as has been outlined in Appendix C of \cite{buchert:dust}, based on \cite{yodzis}).
A second subtlety arises since the GBC theorem being originally valid for (positive-definite) Riemannian manifolds needs to be extended to (Lorentzian) semi-Riemannian manifolds. This extension has been done independently by Avez \cite{Avez} and Chern \cite{MR155261}.

\subsection{Assumptions}
\label{section:hypothesis}

The model universe we consider is a triplet $({\M},\TmetricL,\TT)$ where $({\M},\TmetricL)$ is a connected spatially oriented and time-oriented Lorentzian manifold of dimension $N+1$. $\TT$ is the dust energy-momentum tensor including the cosmological constant as a source term,
$\TT = \varrho \,\underline{\bm u} \otimes \underline{\bm u} - \Lambda\, \TmetricL$, where $\bm u$ is a geodesic unit future time-like vector field on $\M$ (the $4-$velocity of the fluid), $\underline{\bm u}$ the metric-dual one-form of $\bm u$, and $\varrho$ is a non-negative scalar field on $\M$ (the density of the fluid); we assume the triplet $({\M},\TmetricL,\mathbf{T})$ to satisfy the dust perfect fluid Einstein equation
\begin{equation}
\label{eq:einstein}
\CRicL_{ij}-\frac{\RicciScalL}{2}\CmetricL_{ij}=\varrho \, u_iu_j -\Lambda\, \CmetricL_{ij} \ ,
\end{equation}
where $\TRicL$ and $\RicciScalL$ are, respectively, the Ricci tensor and the Ricci scalar of $({\M},\TmetricL)$. Throughout the paper we use the metric signature ($-1,1,1,1$), and set $8\pi G = 1$, and the speed of light $c=1$.

In all sections, we assume $\M$ to be globally hyperbolic, i.e. there exists a foliation by spacelike smooth Cauchy hypersurfaces $(\Sigma_t : t\in \R)$. Furthermore, we assume that there exists an initial Cauchy hypersurface $\initial\Sigma$ of $\M$ such that $\initial\Sigma \perp {\bm u}$.
In section \ref{section:GBC2} for special cases we will consider the case where $\Sigma$ is closed, i.e. compact and without boundary.

\subsection{Aims and strategy}

This work forms a piece of a larger study that aims both at closing the scalar averaging framework and at understanding if and how the topology of a spacetime satisfying the fluid Einstein equations governs the large-scale average dynamics of the Universe. The present work is a first attempt at using the Gauss-Bonnet-Chern theorem in combination with the scalar averaging framework for self-gravitating matter \cite{buchert:review} in the pursuit of those purposes. Further steps will go beyond fixing the global topology by exploring the dynamics of global topology for cosmological manifolds.

We first recall the $N+1$ formulation of scalar-averaged Einstein equations within a fluid-orthogonal foliation of spacetime. We then study the $N=2$ toy-model in which the Gauss-Bonnet theorem allows us to close the system of averaged equations. We recover and detail results of Magni \cite{Magni}. These are compared with recent mathematical results on empty spacetimes ($\varrho=0$ and $\mathbf C  =0$ with $\mathbf C$ the Weyl tensor). We proceed as follows:
\begin{enumerate}
  \item we recall the scalar-averaged Einstein equations  \cite{buchert:dust,buchert:review} in any dimension within a fluid-orthogonal foliation of spacetime in section \ref{section:averages};
   \item we apply the Gauss-Bonnet-Chern theorem with the aim to close the system of averaged equations. This is done in several steps:
\begin{enumerate}
	\item we apply the Gauss-Bonnet formula to a closed Cauchy surface in the case $N=2$ to recover and generalize results of Magni \cite{Magni}, and a result reminiscent of a theorem by Mess \cite{Mess} in section \ref{section:GBC2}; 
	\item before we go any further, at the end of section \ref{section:GBC2}, we spend some space on discussing the relation between the possibility of a dynamical topology change and black hole formation, being largely inspired by results in the case $N=2$, and including some speculations for their generalization to the case $N=3$;
		\item we then give a quick introduction to Gauss-Bonnet-Chern's formula and Chern's form as well as Avez' Wick rotation method in section \ref{section:Chern};
	\item we then move to the case $N=3$ and show, how the sandwich method and the Wick rotation are applied to obtain a formula that we then express through kinematical and gravitational quantities in section \ref{section:GBC4};
	\item in section \ref{section:Results} we obtain a compact relation between averaged curvature invariants and the averaged source fluctuation that enjoys several formulations (in terms of evolution equations for averaged kinematical invariants, averaged scalar curvature or kinematical backreaction), and we evaluate the result for subcases (spatially compact without boundary spacetimes, silent universe models);
	\item here we also verify these results through explicit derivation in the case of exact solutions as subcases of silent universe models, and discuss the results.
\end{enumerate}
\end{enumerate}

\subsection{Notations}

Tensors in abstract form are written bold, $\mathbf{T}$, while the coefficient of tensors in coordinates are normal font-weight, $\mathrm{T}_{ijk...}^{~~~lmn...}$. Einstein's summation convention is used, Latin indices can refer to summation over space  or spacetime indices depending on context.  
 The canonical volume form on $({\M},\TmetricL)$ is denoted by
$\TVolL=\sqrt{|\det \CmetricL_{ij}|}{\bf d}x_0\wedge \cdots \wedge {\bf d}x_N$. We denote by $ \TmetricR$ the induced metric on the leaves $\Sigma_t$  with induced volume form $\TVolR = \sqrt{|\det \CmetricR_{ij}|}{\bf d}x_1\wedge \cdots \wedge {\bf d}x_N$. The volume form of the boundary $\partial\D$ of a spatial averaging domain $\D\,  \equiv \D_t \subset \Sigma_t$, assumed to be compact with smooth boundary, is denoted by $\mathfrak{w}$ with its components denoted by $w_{ij}$.
Given a symmetric $2-$tensor $\mathbf A$, $\inI(\mathbf A)$, $\inII(\mathbf A)$ and $\inIII(\mathbf A)$ denote the principal scalar invariants of $\mathbf A$. They can be defined as the principal scalar invariants of the matrix $(\mathbf A)^i_{~j} = :A^i_{~j}$ in a local chart:
   \begin{eqnarray}
   \inI(\mathbf{A})&= A^i_{~i} \ ;\nonumber \\
   \inII(\mathbf{A})&=\frac{1}{2} ( \,A^i_{~i}A^j_{~j} - A^i_{~j}A^j_{~i} \,) \ ;\nonumber \\
   \inIII(\mathbf{A})&=\frac{1}{6}(\,A^i_{~i}A^j_{~j}A^k_{~k} + 2A^i_{~j}A^j_{~k}A^k_{~i} - 3A^i_{~i}A^j_{~k}A^k_{~j}\,) \ ,
   \end{eqnarray}
or equivalently as the elementary symmetric polynomials in the eigenvalues of $\mathbf{A}$ of degree 1,2 and 3 respectively.
When $\mathbf{A}$ is not specified, $\inI$, $\inII$  and $\inIII$ refer to scalar invariants of $\mathbf \Theta$, the expansion tensor of the flow-orthogonal foliation $\Sigma_t$.

$\RicciScalL,\TRicL,\TRiemL$ denote respectively the Ricci scalar, the Ricci tensor and the Riemann tensor of $(\M,\TmetricL)$, while $\RicciScalR,\TRicR $ and  $\TRiemR$ denote respectively the Ricci scalar, the Ricci tensor and the Riemann tensor of the leaves $(\Sigma_t,\TmetricR)$. 

\section{Spatial volume-averaging of the $N+1$ Einstein equations}
\label{section:averages}

We give ourselves an $N-$dimensional, flow-orthogonal initial Cauchy hypersurface $\initial\Sigma = \Sigma_{t=t_i}$, where we set $t_i \equiv 0$. Define $\Phi_t: \Sigma_{t=0}\rightarrow \M$, the embedding defined by the flow of the vector field $\bm u$ at proper time $t$, $t \in ]t_*,t^*[$; we set $\Sigma_t := \Phi_t(\Sigma_{t=0})$.
Since $\bm u$ is geodesic and normal to $\initial\Sigma$, the foliation $\Sigma = (\Sigma_t)_{t\in ]t_*,t^*[}$ is normal to $\bm u$.
We parametrize $\M$ through the proper time coordinate $t$ of the fluid with the convention that $t=0$ on $\Sigma_{t=0}$; this way $\bm {\partial}_t = \bm u$.

\subsection{$N+1$ equations}

The Raychaudhuri and Gauss-Codazzi equations, the energy constraint and the second Bianchi identity provide the following evolution and constraint equations:
\begin{eqnarray}
\fl\qquad\qquad
\label{eq:localcontinuity}\partial_t \varrho &=& -\inI \varrho \label{eq:localhcons}  \ ;  \\
\fl\qquad\qquad
\label{eq:localgauss}\partial_t \Theta^i_{~j} &=& -\inI \Theta^i_{~j}-{\CRicR}^i_{~j}+\frac{1}{N-1}\left(\varrho+2\Lambda\right)\delta^i_{~j} \ ;  \\
\fl\qquad\qquad
  	\label{eq:localraychaud} \partial_t\inI &=& -\inI^2+2\inII-\frac{N-2}{N-1} \varrho +  \frac{2}{N-1}\Lambda ~~; \\
\fl\qquad\qquad
  	\partial_t \inII &=& -2\inI\cdot \inII+(\rho+2\Lambda)\inI-\RicciScalR\inI+\Theta_i^{~j}\CRicR_{j}^{~i} =-\varrho \inI+\Theta_i^{~j}\CRicR_j^{~i} \ ;  \label{eq:partialII}\\
\fl\qquad\qquad
	\label{eq:partialIII} \partial_t\inIII&=& -3 \inI\cdot \inIII+ \frac{N-2}{N-1}\left(\rho+2\Lambda \right)\inII-\RicciScalR\inII - \Theta_i^{~j}\Theta_j^{~k}\CRicR_{k}^{~i} + \inI\CRicR_{i}^{~j}\Theta_j^{~i}  \ ;\\
\fl\qquad\qquad
	\frac{1}{2}\RicciScalR&=&\varrho+\Lambda-\inII \ ; \label{hamilton}\\
\fl\qquad\qquad
      \inI_{|| j} &=& \frac{N}{N-1}\sigma^i_{\,\;j || i}  \ ;
\end{eqnarray}
where the shear tensor components $\sigma_{ij}$ form the trace-free part of $\Theta_{ij}$, and the double vertical slash denotes covariant derivative with respect to the spatial metric.

\subsection{Volume-averaged $N+1$ equations} 
\label{averagedeqs}

Recall that $\Phi_t $ is defined by the proper time flow of $\bm u$.
For $\initial\D\subset \initial\Sigma$, a regular compact subset of the initial Cauchy hypersurface $\initial\Sigma$ given \textit{a priori}, we define $\D_t:=\Phi_t(\D_{t=0})$; since $\bm u$ is the $4-$velocity of the fluid, the restmass of $\D_t$ does not depend on $t$.
We confine ourselves to the averaging of scalars over $\D_t \equiv \D$.
		
	The $\D$-averaged value $\langle \psi \rangle_{\D}(t)$ of a scalar function $\psi$ is defined by:
\begin{equation}
	\langle \psi \rangle_{\D}(t):=\frac{\int_{\D} \psi \;\TVolR}{\int_{\D}\;\TVolR}\ .
\end{equation}
In section \ref{section:Results} we shall also employ the averaging brackets for surface integration on the boundary of $\D$, i.e.:
\begin{equation}
	\langle \psi \rangle_{\partial\D}(t):=\frac{\int_{\partial\D} \psi \;\mathfrak{w}}{\int_{\partial\D}\;\mathfrak{w}}\ .
\end{equation}
As this averaged scalar function is time-dependent, there is the following commutation relation \cite{buchertehlers,buchert:dust}, see also \cite{ellisbuchert,covariance}.

For any compact domain $\D$ and any scalar field $\psi$ we have:
\begin{equation}
\label{eq:comrule}
\partial_t\langle \psi  \rangle_{\D}-\langle \partial_t \psi \rangle_{\D}= \langle \psi\inI\rangle_{\D}-\langle \psi\rangle_{\D} \langle \inI \rangle_{\D} \ .
\end{equation}
This rule is the basis for the comparison between averaged quantities of inhomogeneous models and homogeneous models. We have the following useful corollary.\footnote{
	In what follows one should understand $(\partial_t+\inI)^k$ as $k$ iterations of the differential operator $\partial_t+\inI$. For instance, $(\partial_t+\inI)^2 \psi= \partial_{tt}\psi +2\inI\partial_t\psi+\inI^2\psi + \dot \inI \psi$. 
The same remark holds for $(\partial_t+\langle \inI\rangle_\D)^k$.} 

For any scalar functions $\psi_0,\psi_1,\cdots, \psi_n$, if
\begin{equation}
\label{comm2}
\sum_{k=0}^n (\partial_t + \inI)^k \psi_k = 0 \ , 
\end{equation}
then, for any compact domain $\D$, we have
\begin{equation}
\label{comm3}
\sum_{k=0}^n (\partial_t + \langle\inI\rangle_\D)^k \langle\psi_k\rangle_\D = 0 \ .
\end{equation}
Therefore, differential relations between local variables that can be written in the form of the corollary have average equivalents. This is the case for the continuity equation and Raychaudhuri's equation: 
\begin{equation}
\label{av1}			(\partial_t+\langle \inI\rangle_\D)\langle \varrho \rangle_\D =0 \ ;
\end{equation}		
\begin{equation}
\label{av2}			(\partial_t+\langle \inI \rangle_{\D})\langle \inI \rangle_\D= 2\langle \inII \rangle_\D 
			- \frac{N-2}{N-1}\langle\varrho \rangle_\D +\frac{2}{N-1}\Lambda \ .
		\end{equation}		
Define:
	\begin{eqnarray}\label{eq:expcoef}
		a_\D&:=&\left(\frac{\int_{\D}\TVolR}{\int_{\initial\D} \TVolR} \right)^{1/N} \ ;\\
			\label{eq:hubble}
			H_\D&:=&\frac{1}{N}\frac{\dot a_\D}{a_\D} \ .
		\end{eqnarray}
The application of the commutation rule \eref{eq:comrule} to the Raychaudhuri equation \eref{eq:localraychaud}, the conservation equations \eref{eq:localcontinuity}, \eref{eq:partialII}, \eref{eq:partialIII}, and the energy constraint \eref{hamilton} on the domain $\D$ yields the following averaged equations for $N \ne 1$ (for $N=1$, Einstein's equation has to be modified see for instance \cite{Boozer})
\textit{cf.} \cite{Magni}: 
		\begin{equation}
			\partial_t\langle \inII \rangle_{\D}=-\langle \varrho \inI \rangle_\D+\langle \Theta^i_{~j}\CRicR^j_{~i} \rangle_\D+\langle \inII\cdot \inI \rangle_\D-\langle \inII \rangle_\D\langle \inI \rangle_\D \ ;
	\label{rayav}
		 \end{equation}		
\begin{eqnarray}
			\frac{1}{2}\langle \RicciScalR \rangle_\D &=&\langle \varrho \rangle_\D +\Lambda -\langle \inII \rangle_\D
			\ ; \label{hamiltonav}\\
				\partial_t \langle\inIII\rangle_\D &=& 
		 -2 \langle\inI\cdot \inIII\rangle_\D+ \frac{N-2}{N-1}\langle\left(\rho+2\Lambda \right)\inII\rangle_\D-\langle\RicciScalR\inII\rangle_\D \nonumber\\ &&\quad - \langle\Theta_i^{~j}\Theta_j^{~k}\CRicR_{k}^{~i}\rangle_\D + \langle\inI\CRicR_{i}^{~j}\Theta_j^{~i}\rangle_\D  - \langle\inI\rangle_\D \langle\inIII\rangle_\D \ ,
	\label{IIIav}
\end{eqnarray}
		together with \eref{av1} and \eref{av2}.
The commutation rule \eref{eq:comrule} is again used to calculate $\dot a_\D$ and $\ddot a_\D$, which together provide the averaged
expansion and acceleration laws for inhomogeneous matter distributions as well as a the continuity equation and the {\it curvature-fluctuation-coupling} \eref{equ:integrab} that assures integrability of \eref{equ:raych} to obtain \eref{equ:ham} ($N\ne 1$):
		\begin{eqnarray}
	\label{equ:raych}	&N\frac{\ddot a_\D}{a_\D}  =-\frac{N-2}{N-1}\langle \varrho\rangle_\D +\frac{2}{N-1}\Lambda +\CQ_\D \ ; \\
		&\frac{(N-1)N}{2}\frac{\dot a_\D^2}{a_\D^2} =\langle \varrho \rangle_\D-\frac{1}{2} \langle \RicciScalR \rangle_\D+\Lambda-\frac{1}{2}\CQ_\D \ ; \label{equ:ham}\\
			&0=\partial_t\langle \varrho \rangle_\D +N\frac{\dot a_\D}{a_\D}\langle \varrho\rangle_\D \ ;
			\label{equ:cont}
\\
\label{equ:cont}		&0=\frac{1}{a_D^{2N}}\partial_t \left(\CQ_\D a_\D^{2N}\right)+\frac{1}{a_\D^2}\partial_t\left(\langle \RicciScalR\rangle_\D a_\D^2 \right), \label{equ:integrab}
		\end{eqnarray}
where the overdot denotes partial time-derivative, equivalent here to the covariant derivative along proper time, and where the \textit{kinematical backreaction term} $\CQ_\D$ is defined as
		\begin{equation}
		\label{QN}
		\CQ_\D :=2\langle\inII\rangle_\D -\frac{N-1}{N}\langle \inI\rangle_\D^2 \ .
		\end{equation}
It is  interesting to compare the averaged inhomogeneous equations to the homogeneous ones. The above equations can be re-written so that the backreaction terms are interpreted as sources, providing an \textit{effective (perfect fluid) energy-momentum tensor} \cite{buchert:perfectfluid} (here we assume null fundamental pressure):
		 \begin{eqnarray}
		\varrho_{\D}^{\mathrm{eff}}:&=&\langle\varrho \rangle_\D-\frac{1}{2}\CQ_\D-\frac{1}{2}{\mathcal{W}_\D} \ ;\\
		p_\D^{{\mathrm{eff}}}:&=&-\frac{1}{2}\CQ_\D+\frac{N-2}{2N}{\mathcal{W}_\D} \ ; \\
		\mathcal{W}_\D &:=& \langle \RicciScalR \rangle_\D - {6k}_{\initial\D}/a_\D^2 \ ,
		 \end{eqnarray}
where we have introduced a new backreaction variable $\mathcal{W}_\D$ as the deviation of the averaged scalar curvature from a (scale-dependent) constant-curvature model.
With these definitions, the familiar Friedmann equations, written for any dimension, are obtained:
		\begin{eqnarray}
			0&=&\frac{N(N-1)}{2}\left(\frac{\dot a_\D}{a_\D}\right)^2 {+ \frac{3k_{\initial\D}}{a_\D^2}}-\varrho_\D^{\mathrm{eff}}-\Lambda \ ;\\
			0&=&N\frac{\ddot a_\D}{a_\D}+\frac{N-2}{N-1}\varrho_{\CD}^{\mathrm{eff}} +\frac{N}{N-1} p_\D^{\mathrm{eff}}-\frac{2}{N-1}\Lambda \ ; \\
			0&=&\dot \varrho_\D^{\mathrm{eff}}+N\frac{\dot a_\D}{a_\D}\left(\varrho_\D^{\mathrm{eff}}+p_\D^{\mathrm{eff}}\right)\ . \label{effectiveeqs}
		\end{eqnarray}
It has to be noticed that, while the standard dust $\Lambda$FLRW model (Friedmann-Lema\^\i tre-Robertson-Walker model with a cosmological constant $\Lambda$) is closed if supplied with an equation of state, here simply $p=0$, the averaged inhomogeneous dust system written above is not closed.
		There are thus several options: either to construct an equation of state, $p^{\mathrm{eff}}=f(\varrho^{\mathrm{eff}}, a_\D)$, which is dynamical and comprises the physics of inhomogeneities, or to consider an explicit model for the inhomogeneities (see the references in the introduction for these two options), or to find new constraining equations.
		
Before going any further, we notice that equations (\ref{equ:raych}\,-\ref{equ:cont}) can be combined to get the following:
	\begin{eqnarray}
	\label{combi}
		\frac{\partial_{tt} (a_\D^N)}{a_\D^N} &= & \frac{N}{N-1} \frac{\langle \varrho \rangle_{\initial\D}}{a_\D^N} + \frac{2N}{N-1}\Lambda -\langle \RicciScalR \rangle_\D \ ; \ \langle \varrho \rangle_{\initial\D} = const.
	\end{eqnarray}
Closure of the scalar averaging framework can thus be reduced in any dimension $N >1$, to constraining the averaged Ricci scalar on the Cauchy hypersurfaces. This is coherent with the situation in Friedmannian cosmology in which the averaged Ricci scalar follows a simple law $\propto a^{-2}$ in every dimension $N>1$.

\section{Closure by Gauss-Bonnet: the $2+1$ equations}
\label{section:GBC2}

Gravity in dimension $2+1$ has many physical limitations; it lacks important dynamical aspects of $3+1$ gravity. It is, however, a toy-model from which one can expect to extract qualitative properties that still apply to higher dimensions (see for instance \cite{carlip} for $2+1$ quantum gravity). Indeed, $2+1$ gravity is so simple that one can compute almost everything and quantify qualitative facts which ease extrapolation to higher dimensions. In particular, the 
Gauss-Bonnet theorem works very well in this context as shown by Magni \cite{Magni}. Finally, we can notice that the averaged formulae we actually use do not employ the full set of Einstein's equations, but only a subset compatible with a weaker version of Einstein's equations that is better suited for lower-dimensional considerations such as those exemplified in \cite{jackiwlowgravity,romero,Boozer}.

As noticed at the end of the preceding section, any constraint on the averaged Ricci scalar on the Cauchy hypersurface is sufficient to close the scalar averaged equations;  the Gauss-Bonnet formula gives us such a constraint.
The $2+1$ system is thus solved completely for closed spacetimes. Below, we provide the complete solution, extending the results given by Magni \cite{Magni}.

\subsection{Application of the Gauss-Bonnet formula and complete calculations}

The averaged equations, specialized to the case of $N=2$ on a two-dimensional spatial domain $\D$ (returning here to the averaged scalar curvature as the variable), are:
\begin{eqnarray}
\frac{\ddot a_\D}{a_\D}&=&\frac{\CQ_\D}{2} +  \Lambda \ ; \label{avA} \\
\frac{\dot a_\D^2}{a_\D^2}&=&\langle \varrho \rangle_\D - \frac{1}{2} \langle \RicciScalR \rangle_\D-\frac{1}{2}\CQ_\D+\Lambda \ ;  \label{avC} \\
0&=&\partial_t \langle \varrho\rangle_\D + 2\frac{\dot a_\D}{a_\D} \langle \varrho \rangle_\D \ ;  \\
0&=& \frac{1}{a_\D^4}\partial_t \left(\CQ_\D a_\D^4\right)+\frac{1}{a_\D^2}\partial_t\left(\langle \RicciScalR\rangle_\D a_\D^2 \right) . \label{avD}
\label{equ:eqn2D}
\end{eqnarray}
In this case, the Gauss curvature $\kappa$ and the scalar curvature agree up to a  constant factor 
(the total curvature $\GBC$ will be defined for the general case in section \ref{GBCformulae}):
	\begin{equation}
		2\pi\,\GBC=\kappa \,\TVolR\;=\;\frac{1}{2}{\cal R}\,\TVolR \ .
	\end{equation}
	The Gauss-Bonnet theorem applied to the domain $\D$ provides:
\begin{equation}
\label{appGBC2}
\langle \RicciScalR \rangle_\D V_\D + 2\int_{\partial\D}v_{i;j}v^jn^i =4 \pi \chi(\D) \ ,
\end{equation}
with $\mathbf v$ the unit tangent vector  of an oriented parametrization of the boundary of $\D$, $\mathbf n$ the outgoing normal vector to $\D$, $v_{i;j}v^jn^i =: k_g$ the extrinsic curvature of the boundary in components, and $\chi$ the Euler characteristic.
In the special case $\D=\Sigma $ (the whole Cauchy hypersurface) with the hypothesis that $\Sigma$ is \textit{compact and without boundary}, it instead provides:
\begin{equation}\label{eq:GBC2D_compact}
\langle \RicciScalR \rangle_{\Sigma}=\frac{4\pi\chi(\Sigma)}{V_\Sigma}\ ; \ V_\Sigma = V_{\initial\Sigma} a_{\Sigma}^2 \ ,
\end{equation}
i.e. the averaged scalar curvature obeys a conservation law, $(\langle \RicciScalR \rangle_{\Sigma} V_\Sigma\dot{)} = 0$. If $\Sigma$ is geodesically complete (instead of compact) without boundary with finite volume and finite total curvature, from Huber's theorem there exists a compact surface without boundary $\overline\Sigma$ and points $(p_i :~ i = 1 \cdots s)$, such that the punctured surface $ \overline\Sigma\setminus\{p_1,\cdots,p_s\}$ is diffeomorphic to $\Sigma$ and we have:
\begin{equation}\label{eq:GBC2D_complete}
\langle \RicciScalR \rangle_{\Sigma}=\frac{4\pi}{V_\Sigma}\left(\chi(\overline\Sigma)-\sum_{i=1}^{s}\kappa_i\right)  \ ; \ V_\Sigma = V_{\initial\Sigma} a_{\Sigma}^2 \ ,
\end{equation}
for  some $\kappa_i\geq 1$ depending of the asymptotic behaviour of the metric of $\Sigma$.
The interested reader can find a statement of Huber's theorem in \cite{Troyanov} and a complete exposition in \cite{Huber}.\\

	Setting the dimensionless volume $\zeta_\D:=\frac{V_\D}{V_{\initial\D}}=a_\D^2$, the entire system can be solved with the convention $t=0$, a time at which $a_{\initial\D}=1$:
	\begin{equation}\fl\qquad
	\ddot\zeta_\D-4\Lambda \zeta_\D = 2 \frac{M(\D)-2\pi\chi(\D)}{V_{\mathbf i}}+\int_{\partial\D} k_g \ \  ; \ \zeta_\D(0)= 1 \ ; \ \dot \zeta_\D(0)= 4 H_{\mathbf i} \ ,
	\end{equation}
	where $M(\D)$ is the total rest mass of the domain $\D$. We consider the situation in which we assume $\D =:\Sigma$ to be closed without boundary. Then we have: 
	\begin{equation}\label{equ:systemz}\fl\qquad
		\ddot\zeta_\Sigma-4\Lambda \zeta_\Sigma= 2\frac{M(\Sigma)-2\pi\chi(\Sigma)}{V_{\mathbf i}} \ \ ; \ \zeta_\Sigma(0)= 1 \ ; \ \dot \zeta_\Sigma(0)= 4 H_{\mathbf i} \ .
	\end{equation}
	We accordingly denote for simplicity $M\equiv M(\Sigma)$, $V_{\mathbf i} \equiv V_{\initial\Sigma}$, $H_{\mathbf i}\equiv H_{\initial\Sigma}$.
From equation \eref{avD}, with $\CQ_{\mathbf i} \equiv \CQ_{\Sigma|t=0}$, one has $
	 \CQ_\Sigma = \CQ_{\mathbf i} a_\Sigma^{-4}$.
Then, using $\ddot a_\Sigma = \ddot \zeta_\Sigma / a_\Sigma - (\dot a_\Sigma)^2 / a_\Sigma $ together with the system (\ref{avA}\,-\ref{avD}) and \eref{equ:systemz} one gets 
$
\CQ_{\mathbf i} =  2\frac{M-2\pi\chi}{V_{\mathbf i}}+2\Lambda-8H_{\mathbf i}^2  ,
$
and, thus  
\begin{equation}
\CQ_\Sigma =  \left(2\frac{M-2\pi\chi}{V_{\mathbf i}}+2\Lambda-8H_{\mathbf i}^2\right) a_\Sigma^{-4} \ .
\end{equation}
We also define $\tau = 1 / (2\sqrt{|\Lambda|})$ for $\Lambda\neq 0$.\footnote{We thank Matthieu Chatelain for fruitful interaction on the following families of solutions within his Master internship report.}

	\subsubsection{Case $\Lambda<0$.}
	The solution of the system (\ref{avA}\,-\ref{avD}) reads:
	\begin{eqnarray}
a_\Sigma (t)&=&\sqrt{\left(1+\frac{M-2\pi\chi}{2V_{\mathbf i} \Lambda}\right) \cos({t/\tau})+{4\tau H_{\mathbf i}}\sin({t/\tau})-\frac{M - 2\pi\chi}{2 V_{\mathbf i} \Lambda}} \ .
\end{eqnarray}
With $A := 1+\frac{M-2\pi\chi}{2V_{\mathbf i} \Lambda}$, $B:=4\tau H_{\mathbf i}$ and $C: =-\frac{M - 2\pi\chi}{2 V_{\mathbf i} \Lambda}$, we define the discriminant, $\Delta := C^2 -(A^2+B^2)$. We have the following cases:
	 
\begin{itemize}
    \item If $\Delta>0$, then $\zeta_\Sigma$ never vanishes and, since $a_\Sigma(0)=1$ is positive, $a_\Sigma$ is defined on $\mathbb{R}$. Since $A=1-C$, we have $C>1/2>0$, thus, $2\pi\chi>M\geq 0$ and, thus, $\Sigma$ is a topological sphere $\Sigma\simeq \mathbb S^2$. A subcase of interest is the equilibrium case where $H_{\mathbf i}=0$ and $M = 4\pi-4 V_{\mathbf i} \Lambda$, so that the scale factor is constant without homogeneity assumption.
    
    \item If $\Delta=0$, then $t_\pm = \tau \left(\arctan{(B/A)}+(2k+1)\pi\pm\pi\right)$, with  $k\in \mathbb Z$, such that $t_-<0<t_+$.  Then, $a_\Sigma$ is defined on $] t_-  ; t_+ [ $. Again, $\chi>0$ and $\Sigma$ is a topological sphere.
    
    \item If $\Delta <0$, we have three subcases: 
    \begin{itemize}
     \item  If $B^2\neq C^2$, define $T_{\pm}$ to be $\tau\arctan\left(\frac{AB\pm C\sqrt{|\Delta|}}{C^2-B^2}\right)+2k_{\pm}\tau\pi$, if $AC \mp B\sqrt{|\Delta|}\geq 0$, and  $\tau\arctan\left(\frac{AB\pm C\sqrt{|\Delta|}}{C^2-B^2}\right)+(2k_{\pm}+1)\tau\pi$ otherwise, with $k_{\pm}$ such that $T_{\pm} \in ]-\tau\pi,\tau\pi[$. Setting $t_{-} = \min(T_-,T_+)$ and $t_+ = \max(T_-,T_+)$, $a_\Sigma$ is defined on $]t_-,t_+[$.
     \item If $B^2= C^2\neq 0 $, define $T_{\pm}$ to be $\tau\arctan \frac{B}{A} \pm \tau\arctan \frac{\sqrt{|\Delta|}}{C}+2k_{\pm}\tau\pi$, if $BC> 0$,  and $\tau\arctan \frac{B}{A} \pm \tau\arctan \frac{\sqrt{|\Delta|}}{C}+(2k_{\pm}+1)\tau\pi$ otherwise, with $k_{\pm}$ such that $T_{\pm} \in ]-\tau\pi,\tau\pi[$. Setting $t_{-} = \min(T_-,T_+)$ and $t_+=\max(T_-,T_+)$, $a_\Sigma$ is defined on $]t_-,t_+[$.
     \item If $B=C=0$, then $a_\Sigma$ is defined on $]-\tau\pi/2,\tau\pi/2[$.
    \end{itemize}
\end{itemize}

	\subsubsection{Case $\Lambda=0$.}
	The solution of the system (\ref{avA}\,-\ref{avD}) reads:
		\begin{eqnarray}
		a_\Sigma (t)&=&\sqrt{\frac{M-2\pi\chi}{V_{\mathbf i}} t^2+4H_{\mathbf i} t+1} \ . 
	         \end{eqnarray}

Defining the discriminant $\displaystyle \Delta := 16H^2_{\mathbf i}-4\frac{M-2\pi\chi}{V_{\mathbf i}}$, the solution above is defined on the following time range:
\begin{itemize}
\item If $M< 2\pi \chi$ (in particular, $\Sigma$ is a topological sphere), then  $\Delta>0$ and $t~\in~\left]\frac{V_{\mathrm i}}{2}\frac{4H_{\mathrm i}-\sqrt\Delta}{2\pi\chi-M},\frac{V_{\mathrm i}}{2}\frac{4H_{\mathrm i}+\sqrt\Delta}{2\pi\chi-M}\right[$.
\item If $M> 2\pi \chi$ (which is always the case if $\Sigma$ is a higher genus surface), then define $t_{\pm}= \frac{V_{\mathrm i}}{2}\frac{-4H_{\mathrm i}\pm\sqrt{|\Delta|}}{M-2\pi\chi}$. We have the following subcases:
\begin{itemize}
 \item if $\Delta<0$, then $t\in \R$.
 \item if $\Delta\geq 0$ and $t_+<0$, then $t\in ]t_+,+\infty[$.
 \item if $\Delta\geq 0$ and $t_->0$, then $t\in ]-\infty,t_-[$.
 \end{itemize}
  \item  If $M=2\pi \chi$ (in particular, $\Sigma$ is a topological sphere or a torus with $M=0$), then we have the following subcases:
    \begin{itemize}
     \item if $H_{\mathrm i}<0$, then $t\in ]-\infty,-1/4H_{\mathrm i}[$.
     \item if $ H_{\mathrm i} =0$, then  $t\in \R$.
     \item  if $H_{\mathrm i}>0$, then $t\in ]-1/4H_{\mathrm i},+\infty[$.        
     \end{itemize}
\end{itemize}

The first case, $M<2\pi\chi$, is topology-driven with an negative effective total mass and both an initial and a final singularity, while the second family, $M>2\pi\chi$, is mass-driven with either an initial or a final singularity.
The critical cases $M =2\pi\chi$, where the mass exactly compensates the topology term, present the same features: two mutually symmetric solutions featuring either an initial or a final singularity, with another critical situation leading to a stationary model universe.

	\subsubsection{Case $\Lambda>0$.}
	The solution of the system (\ref{avA}\,-\ref{avD}) reads:\\
	\begin{eqnarray}
	\fl\qquad
a_\Sigma(t)=\sqrt{\left(1+\frac{M-2\pi\chi}{2V_{\mathrm i}\Lambda}\right)\cosh(t/\tau) + 4\tau H_{\mathrm i}\sinh(t/\tau) -\frac{M-2\pi\chi}{2V_{\mathrm i}\Lambda}} \ . 
	\end{eqnarray} 

Define $A: = 1+\frac{M-2\pi\chi}{2V_{\mathrm i}\Lambda}$, $B: = 4H_{\mathrm i}\tau$ and $C:= -\frac{M-2\pi\chi}{2V_{\mathrm i}\Lambda}$ and the discriminant $\Delta:=C^2+B^2-A^2$. The ranges of times for which the above formula is valid are described below.
\begin{itemize}
 \item Case $A+B = 1+\frac{M-2\pi\chi}{2V_{\mathrm i}\Lambda}+4H_{\mathrm i}\tau\neq 0$:
 \begin{itemize}
  \item If $\Delta<0$, the solution $\zeta$ does not change sign and $\zeta(0)=1>0$; hence.  $t\in\R$.
  \item If $\Delta=0$, then $t\in ]\tau\ln(\max(C/(A+B),0)), +\infty[$ with the convention $\ln(0)=-\infty$.
  \item If $\Delta>0$, we denote $r_\pm := \frac{-C\pm \sqrt\Delta}{A+B}$; several cases occur depending on whether both, only one or none of $r_\pm$ is positive. The signs of $A+B$, $B^2-A^2$ and $C$ determine each situation, we recall that $A+C=1$: 
  \end{itemize}
\vspace{5pt}  
  
  \[  \begin{array}{cc|cc}
 A+B & B^2-A^2 & C <0   & C >0   \\ \hline &&&\\
 +  & \geq 0  &\multicolumn{2}{c}{\qquad 1>r_+>0>r_-} \\&&&\\
 + & <0 & r_\pm >0 & r_\pm<0 \\&&&\\
 - & \geq 0 &  \multicolumn{2}{c}{\qquad 1>r_->0>r_+}\\&&&\\
 - & <0 & \mathrm{Does~not~exists} & r_\pm > 0
 \end{array}\]

$\qquad$The ranges of times follow:
\vspace{5pt}  
     \[ \fl \begin{array}{cc|cc}     
 A+B & B^2-A^2 & C <0   & C >0  \\ \hline&&& \\ 
 +  & \geq 0  & ] \tau\ln\left( r_+\right) \, ;\, + \infty [ & ] \tau\ln\left( r_+\right) \, ;\, + \infty [ \\ &&&\\ 
 + &<0 & ] -\infty \, ; \, \tau\ln\left( r_-\right) [ \ \mathrm{or} \ ] \tau\ln\left( r_+\right) \, ;\, + \infty [ & \mathbb{R} \\ &&&\\ 
 - & \geq 0 & ]- \infty \, ; \, \tau\ln\left( r_-\right) [ & ]- \infty \, ; \, \tau\ln\left( r_-\right) [\\ &&&\\ 
 - & <0 & \mathrm{Does~not~exists}  & ] \tau\ln\left( r_+\right) \, ; \, \tau\ln\left( r_-\right)  [  \\&&&\\  
 \end{array}\]
 \vspace{10pt}
(The case $[\ A+B>0$, $B^2-A^2<0$, $C<0 \ ]$ presents two subcases depending on whether $r_->1$ or $r_+<1$.)
 \item Case $A+B=0$, with $r:= \frac{B}{1+B}=\frac{4H_{\mathrm i}}{1+4H_{\mathrm i}}$:
  \[  \begin{array}{r|l}
 B<-1  &  t\in]-\infty,\tau \ln(r)[   \\ \hline
 -1 \leq  B \leq 0   & t \in \R \\ \hline 
 B>0&t\in ]\tau\ln(r),+\infty[ \\ \hline
 \end{array}\]
\end{itemize}

\subsection{Discussion}
\label{sec:2Ddiscussion}
The following discussion provides known material as well as some speculative thoughts we feel to be relevant for the transition from the $2+1$ toy model to the physical $3+1$ case. The reader who does not want to interrupt the logic of the paper may directly move to section \ref{section:Chern}.

Two ingredients allow us to close the system (\ref{equ:raych}\,-\ref{equ:cont}):
	\begin{enumerate}
	\item based on the combination (\ref{combi}) of equations (\ref{equ:raych}\,-\ref{equ:cont}) featuring the only unknown $\langle \RicciScalR \rangle_\D$,
	 for any $\D$, any assumption on the average scalar curvature leads to closure; 
	\item the Gauss-Bonnet formula providing a topological constraint on $\langle \RicciScalR \rangle_\Sigma$ when $\Sigma$ is closed without boundary.
	\end{enumerate}
	The first is valid in any dimension $N \geq 2$ and any domain, while the second is unfortunately specific to dimension $2+1$. 
	For homogeneous spacetimes $ \RicciScalR  \propto a^{-2}$, and also the boundary term in equation (\ref{appGBC2}) could lead to other volume-dependent terms. Indeed, one could have boundary terms associated to the event horizon of a black hole, say a Schwarzschild black hole, thus constant volume and constant extrinsic curvature, hence inducing a term $\propto a_\D^{-N}$. One could also have a boundary piece `representative' of the average behaviour, where the volume would be $\propto a_\D^{N-p}$, and where $p$ is the codimension of $\partial \D$, and where the extrinsic curvature would be $\propto a_\D^{-1}$ (assuming that the `effective radii' of the boundary are proportional to $a_\D$), giving rise to a term $\propto a_\D^{-(p+1)}$ (for another relevant discussion of curvature invariants of black hole horizons see \cite{Gregoris}). In this discussion we allow the codimension of boundary pieces to be greater than 1. Such `small' boundary pieces can for instance arise as relatively open subsets of the causal boundary; extreme BTZ-black holes, which are considered below, are codimension 2 subsets of the causal boundary, and the initial singularity of spacetimes considered in subsection \ref{sec:qual_top} may have codimension up to $N$.		
Finally, we would be tempted to extrapolate these findings to the existence of a topological or geometrical quantity that would replace the Euler characteristic in higher dimensions, postulating a GBC-type relation of the form: 
\begin{equation}
	\label{scaling}
	 \langle \RicciScalR \rangle_\D = \frac{\alpha_{\initial\D}}{a_\D^N}+\frac{\beta_{\initial\D}}{a_\D^{N-1}}+\cdots \ ,
\end{equation}	   
with constants $\alpha_{\initial\D}, \beta_{\initial\D}, \cdots$.
Such solutions are related to so-called \textit{scaling solutions} that were used to close the system of average equations in many works, \textit{cf.} references in the introduction, or to model curvature evolution to explain observational data without dark energy, e.g.~\cite{celiaSN}, or in perturbation theories. Indeed, a closure by the global stationarity assumption $\ddot a_\Sigma = 0$, yields solutions of the form (\ref{scaling}) for $N=3$ \cite{buchert:global}.

A further possibility that might put a dynamical constraint on $\langle \RicciScalR \rangle_\CD$ for $N+1$ 
is a change in the topology of the Cauchy hypersurface, as it would change $\chi$ in the case $2+1$. To clarify what is meant by topology change, we are aware that Geroch's theorem \cite{Geroch} for globally hyperbolic spacetimes states that the topology of the Cauchy hypersurfaces does not change. On the one hand, the foliation given by Geroch seems not coherent with the intuition we have in the instance of  black hole  formation. Indeed, we imagine a once formed singularity as part of our spacelike leaf, while Geroch's foliation would avoid the singularity and 
would bend the Cauchy hypersurfaces in such a way that even for an arbitrary future leaf there will be a piece of this leaf lying in the past of the singularity. It may appear that a more physically realistic spacelike foliation that would hit a black hole singularity would be cut along the event horizon, so that its future domain of dependence is still the whole model universe outside the event horizon, but such that the past domain of dependence may not contain pieces of the model universe; for instance, there may exist worldlines entering the event horizon in the past of a given leaf, but never intersecting the said leaf. On the other hand, one could argue that no observer actually sees the event horizon and the singularity formation, thus a more causally realistic picture would then be Geroch's. These two viewpoints are actually identical asymptotically for simple black holes such as Schwarzschild's. Indeed, if one considers a leaf given by Geroch's theorem in the far future, the leaf would look similar to a hyperbolic cusp starting to bend violently toward the past in the vicinity of the event horizon. Therefore, at future time infinity, Geroch's foliation tends toward a stationary foliation of the exterior of a black hole.

In any case, a leaf of such a punctured foliation would still contain all the future information a leaf of a Geroch foliation contains, but would allow for topology change. In the case of a Schwarzschild black hole one would have a puncture.
The picture described above is at best valid for Schwarzschild black holes and remains too naive to extrapolate it to more general black hole solutions such as Kerr, Kerr-Newman or Hayward  black holes. Indeed, frame dragging, ring-like naked singularities, among other more complicated phenomena, are expected to  make this picture more involved.
We also might think of other `fantasy' processes that can change the topology like random Einstein-Rosen bridges occurring in the formation of non-deterministic or even naked black holes/singularities. In any of these situations, a piece of boundary around a black hole could evolve in a sophisticated manner which would be very different from the simple Schwarzschild picture {(for illustrations see, e.g. \cite{Eric:BH}).}
In the remainder of our discussion, we emphasize the qualitative importance of black holes and topology change for the large-scale average dynamics. 

\subsubsection{Qualitative influence of topology on spacetime dynamics}
\label{sec:qual_top}

The preserved quantity is of topological origin in the $2+1$ context: the Euler characteristic is a topological invariant. This topological quantity is given by the following formula for finitely-punctured, compact and oriented surfaces: 
  \begin{equation}
  \chi({\Sigma_t}) = 2-2g-s \ ,
  \end{equation}
where $g$ is the genus (the number of handles) {of the leafs $\Sigma_t$}, and $s$ is the number of punctures (the number of `missing' points). More generally, by Richards' theorem \cite{MR143186}, a topological surface can be constructed from a sphere by adding handles, glueing Klein bottles (possibly infinitely many) and puncturing it.
  
We would like to compare the picture that $\chi$ may govern the large-scale behaviour with the situation where $\TRiemL= \mathbf{0}$. Every Lorentzian manifold satisfying $\TRiemL = \mathbf{0}$ everywhere is locally isometric to Minkowski space (i.e. there exists a neighborhood of each point which embeds  isometrically into Minkowski), but this is not true globally. A theorem of Mess \cite{Mess}, since then generalized by Bonsante and Benedetti \cite{bonsante_flat_2003,MR2499272}, Barbot \cite{barbot_globally_2004}, Bonsante and Seppi \cite{MR3493421}, and Brunswic \cite{brunswic:hal-01401821}, `classifies' Cauchy-compact maximal globally hyperbolic locally Minkowski spacetimes with singularities such as massive particles or BTZ-like black holes (see next paragraph). For instance, spacetimes without singularities are all quotients of some convex polyhedron of Minkowski space by a discrete subgroup of the Poincar\'e group. Three representative situations can be distinguished: 
\begin{itemize}
\item  sporadic cases we will neglect,
\item flat periodic spacetimes or translation spacetimes: 
$$ \M= \mathbb T^p \times \R^{N-p}\times \R\ ,\quad  \d s^2_\M=\d s^2_{\mathbb T^p\times \R^{N-p}} -\d t^2 \;,$$
where $\mathbb T^p$ is an $p-$dimensional flat torus for some $p\in [ 0,N]\cap \mathbb N $ $\,$,
\item  and affine deformations (in a sense we shall not discuss) of Milne-like spacetimes: 
	 $$ \M= \Sigma_0 \times \, ]0,+\infty[ \ , \quad \d s^2_\M = t^2 \d s^2_{\Sigma_0}-\d t^2 \;,$$
where {$\Sigma_0$} is a geodesically complete $N-$dimensional Riemannian manifold of constant curvature~$-1$. 
\end{itemize}
Translation spacetimes are stationary, while Milne-like spacetimes are expanding (up to time reversal).
We put into perspective the following.

\begin{theo}[Barbot]
Up to finite coverings and linear twisted products, any $N+1$--dimensional maximal Cauchy-complete globally hyperbolic locally Minkowski spacetime is either sporadic, a translation spacetime, or a twisted product of a Euclidean torus by a deformation of a Milne-like spacetime.
\end{theo}
The expression `finite coverings' means that one can have finite-type symmetries as in a crystal, and `linear twisted products' means one can have large-scale periodicity which is slightly more sophisticated than translation spacetimes.

The topology of the Cauchy hypersurface, up to a few elementary cases, is enough to determine whether the spacetime is dynamic (expanding or contracting) or stationary. This fact can be seen in the $2+1$ computations: for $\varrho=0$, the Euler characteristic solely determines the large-scale behaviour of the volume scale factor $a_\D(t)$. Furthermore, the existence of a big bang or a big crunch is forced by the topology. These two results thus extend to the cases $N+1$ qualitatively (even though the Euler characteristic is not a sufficient criterion anymore), as long as the Riemann tensor vanishes. We summarize these results  by the following qualitative statement: 
   \textit{The more complex the topology of $\Sigma_t$, the more expanding or contracting}. Section~\ref{higherdim} discusses ways to give more precise meaning to this statement. We would expect this phenomenon to be stable upon small metric deformations. 

\subsubsection{Topology and black holes in $2+1$}

A puncture in a finite-type Cauchy hypersurface is naturally interpreted as a singularity, i.e. a black/white hole. Thus, each black/white hole intuitively accounts for $1$ in the $s-$term and $s$ would thus be the number of black/white holes.
This intuition is confirmed in the $2+1$ locally Minkowski case. Barbot, Bonsante and Schlenker \cite{Particules_1,Particules_2} classified $2+1$ geometrical singularities among which two types of black/white holes
are of particular interest. These are variations of the Ba\~nados-Teitelboim-Zanelli (BTZ) Anti-de Sitter example \cite{1992PhRvL691849B}.
Brunswic \cite{brunswic:hal-01317447} extended Mess' theorem to $2+1$ flat spacetimes with so-called extreme BTZ black/white holes. An extreme BTZ black hole is a singularity modeled on $(\R^3,\TmetricL_{BTZ})$, where $\d s^2_{BTZ} = 2\d t\d r+ \d r^2+ r^2\d \theta^2$, using cylindrical coordinates of $\R^3$. The lines $\{r=r_0,\theta=\theta_0\}$ are lightlike; furthermore, the line $\Delta := \{r=0\}$ is singular and behaves as the horizon of a black hole since for any event $p\in \Delta$, with $J^+(p)$ the causal future of $p$, we have $J^+(p)\subset \Delta$.
He shows that one can ``fill'' every puncture of a parabolic holonomy with an extreme BTZ black/white hole in an absolutely maximal spacetime in a unique way. 

\begin{theo}[Brunswic]
Let $\M$ be an absolutely maximal flat spacetime and let {$\Sigma_0$} be a Cauchy surface of $\M$. Then, $\M$ admits a unique extension $\overline \M$ among absolutely maximal flat spacetimes with extreme BTZ-like singularities. Furthermore, the complement of the extreme BTZ-like singularities in $\overline \M$ is isometric to $\M$, and the punctures of {$\Sigma_0$} of parabolic holonomy correspond bijectively to extreme BTZ-like singularities.
\end{theo}
It was already known that punctures with hyperbolic holonomy correspond to BTZ black/white holes; hence, every puncture in a maximal flat Cauchy-complete surface can be filled by a BTZ black/white hole, either extreme or not.  We have: 
\begin{equation}\label{eq:chi_BH}
   \chi = 2-2g - \mathrm{Number~of~(extreme)~BTZ~black/white~holes}\ .
\end{equation}
The genus does not change by the process of black hole creation or white hole destruction.
Finally, we infer that black holes and white holes have a quantitative effect on the large-scale dynamics of a spacetime in dimension $2+1$. We extrapolate such an effect to dimension $N+1$. 
To conclude, we have:
\begin{itemize}
\item a quantitative link between topology (via the Euler characteristic) and large-scale dynamics of a $2+1$ dust spacetime given by the Gauss-Bonnet formula;
\item a quantitative relation between the number of black/white holes and the Euler characteristic in $2+1$ empty spacetimes given by Barbot and Brunswic theorems;
\item a qualitative link between multi-connectedness and large-scale dynamics for $N+1$ empty spacetimes given by Mess-like theorems. 
\end{itemize}
We are thus invited to think that black/white holes have a determining effect on the large-scale average dynamics via their influence of the dynamics of the topology. The above discussion leads for instance to a phase transition in dimension $2+1$: the formation of more than three black holes implies a large-scale contribution to expansion or contraction. This in turn can give rise to a dark energy-like effect in terms of topological volume acceleration, $\ddot a_\CD > 0$.
(For related investigations see, e.g. \cite{DEtopology1,DEtopology2}.\footnote{The term `topological acceleration' has been introduced earlier in \cite{boudtopacc1,boudtopacc2,jantopacc} but referring to a residual acceleration due to the global topology of a manifold; see the critical discussion and corrections by Steiner \cite[\textit{cf.} Refs. 36, 37 therein]{frankBH}. It may  happen that this residual acceleration averages out to zero globally; the topological acceleration generated through local topology change, as defined here, always generates a global effect.})

\subsubsection{Topology, curvature and black holes in higher dimensions} 
\label{higherdim}

The relationship given by equations \eref{eq:GBC2D_complete} and \eref{eq:chi_BH}
linking the average of the scalar curvature to topology and black holes does not generalize to higher dimensions. There are strong mathematical and physical reasons preventing a na\"\i ve generalization. 

First, scalar curvature is very loosely related to topology. Although
there are many strong results constraining the topology of a manifold that admits a strictly positive scalar curvature, the physical reality of inhomogeneous matter distributions implies that, generically, the scalar curvature is somewhere negative. Indeed, on a compact manifold of dimension of at least $3$, every function that is negative at at least one point is the scalar curvature of some Riemannian metric \cite{bergery,kazdan}. 
Even worse, on any manifold of dimension at least 3, one can construct a Riemannian metric of negative Ricci tensor; hence, there exists a negatively curved three-sphere with non-constant sectional curvature (the scalar curvature being the average of the sectional curvatures of the manifold at all the tangent 2-planes at a given point) \cite{lohkamp}. As a consequence, one could have a spherical spacetime that is, at least for some period of time, perturbatively close to a constant negative curvature spacetime. 

Second, as a measure for topological complexity, the Euler characteristic fails in the case of three-manifolds. It determines the topology of compact surfaces (and to some extent the topology of 
finitely-punctured compact surfaces) in dimension $2+1$, but it is not true anymore for higher-dimensional manifolds. For instance, the Euler characteristic is zero for odd-dimensional closed manifolds. The coarse topological invariant provided by the Euler characteristic should thus be replaced by a stronger one. Several candidates could play a role coherent with the $N+1$ Mess theorem: the number of generators of the first fundamental group $\pi_1(\Sigma)$  \cite{Bredon}, the growth rate of $\pi_1(\Sigma)$ (in the sense of geometric group theory, see \cite{pak} for a nice introduction), Betti numbers  \cite{Bredon}, Reidemeister torsion \cite{Nicolaescu}, etc.

Third, the mass spectrum of extreme BTZ black holes in dimension $2+1$ is a singleton: every black hole considered above has the same mass. Obviously, this does not carry over to dimension $3+1$. One should thus weight the number of black holes by a suitable invariant. Again, there are several candidates: the matter content of a black hole, the area of the event horizon, the strength of the gravitomagnetic field, etc. 
Quantifying the $3+1$ case would open the door to observational determination of the global topology by providing observational measures of the number, the mass function and other characteristics of black holes.

Fourth, the formation of a black hole \textit{a priori} adds a puncture $p \in \Sigma$ to a Cauchy hypersurface $\Sigma$. Such a puncture changes the Euler characteristic the same way in every dimension: $\chi(\Sigma\setminus\{p\})= \chi(\Sigma)-(-1)^N$. 
However, as stated above, it is not clear whether the Euler characteristic is the right topological invariant to consider. The only qualitative evidence so far that topology influences large-scale behaviour is Barbot's generalization of Mess' theorem; in this theorem, the topological ends of a Cauchy hypersurface $\Sigma$ of Cauchy-hyperbolic (expanding) spacetimes are $(N-1)-$dimensional tori $\mathbb T^{N-1}$. Therefore, the qualitative result of Barbot does not give any information on the contribution of punctures. Either the qualitative statement is valid for topological ends such as punctures, or one needs to consider more complicated horizons and singularities such as toroidal horizons and to show they are physically relevant (which is not clear since such horizons are unstable in asymptotically flat spacetimes \cite{hawking,galloway}).

{\it Remark:} In this context, a sustained attempt on investigations of black hole lattices is followed by Clifton and collaborators, see the review
\cite{latticesBH:review} and references therein and \cite{latticesBH}, as well as Fennen and Giulini \cite{fennen2019lie} who investigate black hole packings. Such approaches provide constructions of exact initial conditions of spacetimes with black holes allowing to numerically explore the effect of black holes on large scales. The approach by Fennen and Giulini adds a particular interesting element, since black hole packings have a large number of degrees of freedom to allow construction of realistic all-scale matter distributions.

Fifth, topological considerations are closely related to gauge choices: if one consider a globally hyperbolic manifold $\M$, Geroch's splitting theorem ensures that $\M$ splits into a product $\Sigma\times \R$, and all Cauchy-hypersurfaces are homeomorphic. Topology change in a globally hyperbolic manifold has then to be understood as a change of foliation, thus potentially a change of the domain of dependence, in the end a change of the Lorentzian manifold. Such a change has to be justified carefully. In our discussion on foliations around a forming black hole we argued that in the future both choices (foliations through or avoiding the horizon) are acceptable since they are asymptotically equivalent; the information in the future is also conserved since the future domain of dependence of a leaf stopping at the event horizon of a black hole is still supported by initial data that describe the whole exterior of the black hole.
In the $2+1$ setting, the homeomorphism class of a surface is characterized by its genus as well as the topology of its ends \cite{MR143186}. Punctures are but one kind of `end'; physical phenomena that induce a topological dynamics creating more sophisticated ends are not known to the authors. Similarly, a genus change is highly speculative. One can consider topology change in the future or in the past; a change of topology (hence a change of foliation) toward an initial singularity has to be unique or natural in some sense. Belraouti \cite{MR3704814} constructed an initial singularity limit for constant ($N+1-$dimensional) sectional curvature; moreover he proved that a change of foliation is allowed and does not change the limit as long as the leafs are quasi-convex Cauchy hypersurfaces. The initial singularity is then a well-defined simply-connected metric space, though not a manifold.
Other constructions found in \cite{boudLTB} modify the fundamental group or the number of connected components via glueings and/or inhomogeneities. While these constructions are mathematically correct, the discussion lacks both the description of a physical phenomenon producing glueings and a physical justification of the choice of foliation. A natural choice of foliation is provided by the cosmological time \cite{Cosmologicaltime}, which is compatible with the results of Belraouti, though in a quite different context. This choice would not give the same conclusion regarding topology change toward the initial singularity. Again, this stresses the importance of a physically motivated choice of foliation (\textit{cf.} \cite{buchert:generalfluid}).

\section{The Gauss-Bonnet-Chern theorem for semi-Riemannian manifolds}
\label{section:Chern}

In order to investigate the relevant $3+1$ setting, the main focus of this section is to first remind the reader of some material about the Gauss-Bonnet-Chern (henceforth GBC) formula and its generalizations to semi-Riemannian manifolds. In the whole section, $n=2p$ is an even integer and $(\M, \TmetricL)$ is a $2p-$dimensional semi-Riemannian manifold, thus, it might be Lorentzian or Riemannian depending on the context. We recall that the Euler characteristic is necessarily trivial for any compact without boundary odd-dimensional manifold \cite[pp. 340-358]{Bredon}. 

The original work can be found in Chern's paper \cite{Chern1}, who uses the Cartan formalism in the Riemannian case. Subsequently, Avez \cite{Avez} and Chern \cite{MR155261} generalized the theorem to semi-Riemannian manifolds independently and using very different techniques. We will follow the method of Avez which can be directly extended to manifolds with boundaries. The authors are aware that GBC formulae for semi-Riemannian manifolds satisfying various gravitation/quantum laws are an active thread of research of which we do not produce a bibliography, since its objectives are very different from ours and its results are inapplicable to our purposes.

\subsection{The Riemannian Gauss-Bonnet-Chern formulae}
\label{GBCformulae}

The GBC term (or total curvature $\GBC$) is defined as:
  		\begin{equation}
			\GBC=\frac{(-1)^{p-1}}{2^{3p}\pi^p p! }\CVolL_{a_1,\cdots, a_n}\CVolL_{b_1,\cdots, b_n}\prod_{i=1}^p\CRiemL^{a_{2i-1} a_{2i} b_{2i-1}b_{2i}}\TVolL \ ,
		\end{equation}
		which gives in the $4-$dimensional case:
		\begin{equation}\label{GBC4D}
			\GBC=\frac{-1}{128\pi^2  }\CVolL_{a_0,a_1,a_2, a_3}\CVolL_{b_0,b_1,b_2, b_3}\CRiemL^{a_{0} a_{1} b_{0}b_{1}}\CRiemL^{a_{2} a_{3} b_{2}b_{3}}\TVolL\ .
		\end{equation}
		The following theorem is well-known.
		\begin{theo}[Gauss, Bonnet, Weyl, Allendorfer, Chern \cite{Chern1}]
		Let $\M$ be a Riemannian manifold without boundary, then we have:
		 \begin{equation}
		 \int_{\M} \GBC = \chi(\M)\ .
		 \end{equation}
		\end{theo}
However, the core result of Chern is stronger. Indeed, Chern's simple proof boils down to showing that even though $\GBC$ is not an exact form, $\GBC$ can be lifted to a bundle on which it is exact: the sphere bundle.

\begin{defi}[Sphere bundle] Let $(\M,\TmetricL)$ be a $2p-$dimensional Riemannian manifold. The sphere bundle of $\M$, denoted by $\mathbb S \M$, is the subbundle of the tangent bundle $T\M$ whose elements are the unit vectors.
\end{defi}
  		\begin{theo}[Chern \cite{Chern1}]\label{theo:Chern}
  		Let $(\M,\TmetricL)$ be a $2p-$dimensional Riemannian manifold, 
		and let $\pi: \mathbb S \M \rightarrow \M$ be its sphere bundle, $\pi^*$ denoting the pullback by $\pi$. 
  		Then, there exists a $(2p-1)-$form $\boldsymbol \Pi$ such that 
\begin{equation} 
\pi^* \GBC = {\bf d} \boldsymbol{\Pi} \ .
\end{equation}
Furthermore, $\boldsymbol\Pi$ can be decomposed as follows:
\begin{equation}\label{eq:Pi}
\boldsymbol{\Pi}= \frac{1}{\pi^p}\sum_{k=0}^{p-1} (-1)^k \frac{1}{1\cdot 3\cdots (n-2k-1)k!2^{p+k}} \boldsymbol{\phi}_k \ ,\end{equation}
where, for any local section $\boldsymbol{\ell}:\  \U \subset \M \rightarrow \mathbb S \M$, 
the pullback of $\boldsymbol{\phi}_k$ by $\boldsymbol{\ell}$ can be written:
\begin{equation}
\fl
(\boldsymbol{\ell}^*\boldsymbol{\phi}_{k})_{c_1\cdots c_{2p-1}} = \frac{1}{2^k} \CVolL_{a_{1}\cdots a_{2p}}\CVolL^{b_1\cdots b_{2p} } \ell^{a_1}\prod_{i=2}^{2p-2k}\ell^{a_{i}}_{~;b_i}\prod_{i=2p-2k}^{2p-2} \CRiemL^{a_{i+1}a_{i+2}}_{~~b_{i+1}b_{i+2}} \CVolL_{b_1c_1\cdots c_{2p-1}}\ ,
\end{equation}
with $\TVolL$ the volume form of $(\M,\TmetricL)$.
\end{theo}

$\boldsymbol\Pi$ involves first order derivatives of $\boldsymbol{\ell}$ which is expected, since it is defined on a submanifold of the tangent bundle of $\M$.

To deduce the Gauss-Bonnet-Chern formula, we remark that $\boldsymbol{\phi}_0$ is the (normalized) volume form on the fiber $\boldsymbol{\Pi}^{-1}(p)$, $p \in \M$, induced by the Riemannian metric, and that we choose a singular vector field $\mathbf{v}$ on $\M$ with isolated order $\pm1$ singular points.
One can then apply Stokes' formula on the image of $\M':=\M\setminus \{{\mathbf{v}=\mathbf{0}}\}$ by 
the normed vector field $\boldsymbol{\ell}=\frac{1}{|\mathbf{v}|} \mathbf{v}$. One then obtains:
\begin{equation}
\fl \qquad
\int_{\M}\GBC = \int_{\boldsymbol{\ell}(\M') }\pi^*\GBC = \int_{\partial\boldsymbol{\ell}(\M')}\boldsymbol \Pi =
\sum_{\mathbf{x}\in \{\boldsymbol{\ell}=\mathbf{0}\}} \mathrm{Index}_x(\boldsymbol{\ell})\int_{\mathbb S_{\mathbf{x}}\M} \boldsymbol\phi_0 = \;\mathrm{Index}(\boldsymbol{\ell}) \ .
\end{equation}
By the Hopf-Poincar\'e Theorem \cite{Bredon}, the index of any such vector field $\boldsymbol{\ell}$ equals the Euler characteristic.
                
\subsection{The semi-Riemannian Gauss-Bonnet-Chern formulae}
\label{subsection:GBCSR}

These last theorems do not apply directly to semi-Riemannian manifolds, since Chern makes use of the action of $O(n)$ on tangent spaces which does not preserve the metric tensor on general semi-Riemannian manifolds. We generalize Chern's result using a Wick rotation method, first used by Avez \cite{Avez}; another way based on the Chern-Weyl homomorphism is explained in \cite{MR155261}.

		We give ourselves a semi-Riemannian manifold $(\M,\TmetricL)$ of dimension $2p$ and, as Avez did, notice that all the terms involved in the formula are polynomials in $\TRiemL$,
		$ \TmetricL$, and the coordinates of a vector field. The idea is then to decompose $\TmetricL$ into a 
		positive-definite part $\TmetricL^+$ and a negative-definite
	        part $\TmetricL^-$: $\TmetricL=\TmetricL^++\TmetricL^-$.
		It is possible to deform $\TmetricL$ into $\tilde \TmetricL=\TmetricL^+-\TmetricL^-$ which is positive-definite with explicit expression of the
		integrand along the deformation.

We set that $\TmetricL$ has signature ($k$, $2p-k$), meaning that the matrix $(\TmetricL)_{ij}$ has $2p-k$ positive eigenvalues and $k$ negative ones.
		Given a point $P\in {\M}$, since $(\TmetricL)_{ij}$
		is symmetric at $P$ in any local chart orthonormal at $P$, there is a non-unique decomposition
		of the tangent bundle of $\M$,
		\begin{equation}
		\label{eq:dectangent}
			\displaystyle T{\M}=T^+{\M}\bigoplus^\perp T^-{\M} \ ,
		\end{equation}
		such that $\TmetricL$ induces a
		positive-definite bilinear form on $T^+{\M}$ and a negative-definite on $T^-{\M}$. We can define
		$\mathbf a(x,y)=\TmetricL(\pi^+(x),\pi^+(y))$ and $\mathbf b(x,y)=\TmetricL(\pi^-(x),\pi^-(y))$ with $\pi^+$ and $\pi^-$ the orthogonal projection
		on $T^+{\M}$ and $T^-{\M}$, respectively.
		We then set $\TmetricL^\alpha=\mathbf a -\alpha \mathbf b$ for $\alpha\in \R$, so that $\TmetricL^{-1}=\TmetricL$,\footnote{The notation $\TmetricL^{-1}$ should not be confused with the inverse metric; it refers to $\TmetricL^\alpha$ with $\alpha = -1$.} $\TmetricL^{<0}$ is indefinite, $\TmetricL^{>0}$ is positive-definite,
		and $\TmetricL^0$ is singular. 
		We use the superscript $^{\alpha}$ to denote objects, in particular, $\GBC^\alpha$, associated with the
		metric $\TmetricL^\alpha$. 
		We begin with a key lemma by Avez.
		\begin{lem}[Avez \cite{Avez}]\label{lem:avez} Using the same notations as above,
			there exists a polynomial $\CalA$ with coefficients in $ \CC^\infty({\M},\R)$ depending only on $\mathbf{a}$ and $\mathbf{b}$ such that
			\begin{equation}\label{equ:avez}
				\forall \alpha\in \R^*\ ,\
				\GBC^\alpha=\frac{|\alpha|^{k}}{\alpha^{7p}} \CalA(\alpha)\TVolL^\alpha \ . 
			\end{equation}
		\end{lem}		
		This lemma allows to compute formulae for $\alpha>0$ and extend them for $\alpha<0$ using the analyticity of $\CalA$. For instance, Avez used it to prove an extension of the GBC theorem for compact without boundary semi-Riemannian manifolds. Such manifolds have cyclic time and a further generalization will be needed to obtain a formula applicable to a usual globally hyperbolic setting, which is the subject of section~\ref{section:GBC4}.	

		\begin{theo}[Gauss, Bonnet, Chern, Avez \cite{MR155261,Avez}] Let ${\M}$ be a semi-Riemannian manifold with signature ($k$, $2p-k$), compact without boundary, then
		\begin{equation}
				\int_{{\M}} \GBC=(-1)^{[k/2]}\; \chi(\M) \ .
		\end{equation}			
		\end{theo}
		\textit{Proof:} 
		There are two cases, we begin by the even $k$ one.  Applying the GBC formula for $\alpha>0$, we get:
			$$\int_{\M}\GBC^\alpha=\chi({\M})\ .$$
			One can check that $\TVolL^\alpha = |\alpha|^{k/2} \TVolL$; then, using lemma \ref{lem:avez}, we   
			 have:
			$$\int_{{\M}} \CalA(\alpha) \TVolL \frac{|\alpha|^{3k/2}}{\alpha^{7p}}=  \chi({\M})\ ,$$
			and, thus,
			$$\forall \alpha>0,\quad\alpha^{3k/2 }\int_{{\M}} \CalA(\alpha)\TVolL=\alpha ^{7p}\chi({\M})\ . $$
			If $k$ is not even, the right-hand side is a polynomial in $\alpha$ while the 
			left-hand side has a factor $\sqrt{\alpha}$, therefore, both are zero and the theorem is true. If $k$ is
			even, since the right-hand side is a polynomial in $\alpha$, one can extend the formula to negative 
			alpha: 
			$$\forall \alpha \in\R^*,\quad\alpha^{3k/2}\int_{{\M}} \CalA(\alpha)\TVolL=\alpha^{7p}\chi({\M})\ . $$	
			In particular, for $\alpha=-1$ we obtain: 
			 $$\int_{\M} \GBC = \int_\M \frac{|-1|^{3k/2}}{(-1)^{7p}}\mathcal{A}(-1)\TVolL = (-1)^{3k/2}\int_\M  
			 \frac{(-1)^{3k/2}}{(-1)^{7p}}\mathcal{A}(-1)\TVolL = (-1)^{3k/2}\chi(\M). \quad\Box $$
						
		It is important to extract the general method from the proof above: 
			\begin{enumerate}
			\item finding a decomposition of the tangent bundle as in equation (\ref{eq:dectangent});
			\item considering the twisted metric and giving oneself a unit vector field $\mathbf X$, then cutting the    
			boundary term into a part that can be pulled-back to ${\M}$, the `spatial part', and a part which is purely
			 `angular', identifying the angular part by using the Hopf-Poincar\'e theorem      
			  \cite{Bredon} to obtain the Euler characteristic;
			\item evaluating how the spatial boundary term changes while twisting-back the metric to the semi-Riemannian one, $\alpha$ staying positive;
			\item using the rigidity property of the Avez lemma to extend the formula to negative $\alpha $.
		\end{enumerate}
		The next section is devoted to an adaptation of this scheme in order to obtain, starting from Theorem~\ref{theo:Chern}, new formulae in the $3+1$ averaging context.	
		
\section{Sandwich approach with semi-Riemannian GBC}
\label{section:GBC4}

Considering the results obtained in dimension $2+1$ using the GBC formula and the generalization of the GBC formula to dimension $2p$ and to Lorentzian manifolds we described in the preceding sections, it appears natural to try applying the GBC formula the same way to dimension $3+1$. However, several issues arise. 

First, we applied the GBC formula to a 2D (thus even-dimensional) closed spacelike leaf, while we have now a 3D (thus odd-dimensional) spacelike leaf. In this context, the GBC integrand is identically zero, thus the GBC theorem does not give anything more than the equality $\chi=0$ for closed Cauchy hypersurfaces $\Sigma_t$. One can by-pass this issue by considering a thick spacelike slice $\Sigma_{[t,t+\varepsilon]}$ and by letting $\varepsilon\rightarrow 0$ (so-called sandwich method).
Second, such a thick spacelike slice still has zero Euler characteristic. One cannot expect much topological constraint from this.
Third, one cannot expect gravity to be fully constrained by topology since even in the case of an empty model universe one can still have non-zero Weyl tensor and, thus, a pure curvature dynamics (like gravitational waves). One can have black hole formation even in this setting (see for instance \cite{Christodoulou}).
Fourth, the GBC  integrand is not the Ricci scalar $\RicciScalR$ anymore but a quadratic curvature polynomial.

However, these issues do not prevent us from computing what GBC gives in $3+1$: Chern's theorem \ref{theo:Chern} can be  interpreted as a divergence formula for curvature. Therefore, we still expect that the GBC formula will give us a non-trivial equation in which the Weyl curvature will contribute. Furthermore, the integral nature of the formula is well-suited to give a further non-trivial equation for averaged variables. Indeed, we shall be able to provide such an equation one level further down the hierarchy of a system for scalar averages of $3+1$ Einstein equations, which opens a promising route to a physical closure of this system. We shall argue in \ref{app:closure} that this form makes the closure issue \textit{unique} and more transparent and may pave the way to solve the closure problem, or at least to provide improved closure approximations.
	 
 \subsection{Sandwich method and Wick rotation}
  
This subsection explores a sandwich method in combination with Avez' Wick rotation method in the special case of a $(3+1)-$dimensional spacetime to obtain a new average formula from GBC. 
As before, $(\M,\TmetricL)$ is a $4-$dimensional Lorentzian globally hyperbolic manifold with spacelike leafs $(\Sigma_t,\TmetricR)$ satisfying the Einstein equation for irrotational dust. We write ${\M}= (\R\times\Sigma_t,\TmetricL)$. 
	 
A family of local charts $(t, x_1, x_2, x_3)$ such that $(\boldsymbol{\partial}_{1,2,3}\in T\Sigma_t,\boldsymbol{\partial}_t=\boldsymbol{u})$, where $T\Sigma_t$ denotes the set all vectors $\mathbf{v}\in T\M$ tangent to one of the spacelike leafs $\Sigma_t$, and $\TmetricL=- {\bf d}t\otimes {\bf d}t+\TmetricR$,
allows us to construct a family of semi-Riemannian structures ${\M}^{\alpha}=(\R\times\Sigma_t,\TmetricL^\alpha)$ for $\alpha\in [-1,1]$ with
		\begin{equation}
		\label{overbarg}
		\TmetricL^\alpha=\alpha\,{\mathbf d}t\otimes {\mathbf d}t+\TmetricR\ .
		\end{equation}
		Defined that way, these are the Wick rotations induced by the foliation $\Sigma_t$ (as section \ref{subsection:GBCSR} assumed the choice of such a foliation).
		This family of  symmetric $2-$tensors is singular only for $\alpha=0$, Riemannian for $\alpha>0$ and Lorentzian for $\alpha<0$.
		All objects associated to ${\M}^\alpha$ will be denoted $\star^\alpha $. Our Lorentzian manifold is ${\M}={\M}^{-1}$; the objects associated to $\M^{-1}$ will be denoted without
		any exponent. $\mathbf u^\alpha=|\alpha|^{-1/2}\boldsymbol\partial_t$ is then the unit vector field on $\M^\alpha$ normal to $\Sigma_t$.

Furthermore, since $\boldsymbol \Pi^\alpha$ is a $3-$form and since {${\mathbf u}\perp {\mathbf v}$,} we can write the orthogonal decomposition,
\begin{equation}
\forall \alpha>0,\quad \boldsymbol{u}^*\boldsymbol \Pi^\alpha = \Pi^\alpha_{//}\, \boldsymbol\eta + \Pi^\alpha_\perp \underline{\boldsymbol{u}}\wedge \mathfrak{w} + \boldsymbol\zeta \wedge \underline{\boldsymbol{v}}\wedge \underline{\boldsymbol{u}} \ ,
\end{equation}
where $\TVolR$ is the volume form induced on $\Sigma_t$, $\mathfrak{w}$ (with components denoted by $w_{ij}$) the volume form on the boundary $\partial \D$ induced by $\TVolR$, and $\boldsymbol \zeta$ is some $1-$form.
Notice that the formulae for ${\mathbf u}^*\boldsymbol\phi_k$ make sense whether the metric $\TmetricL$ is Riemannian or Lorentzian and whether the vector $\mathbf u$ is unit or not. We thus extend the formula for 
${\mathbf u}^*\boldsymbol \Pi^\alpha$ for all $\alpha$. Then, taking any $\alpha>0$ and writing $\GBC^\alpha = \mathrm{G}^{\alpha}\TVolL$, we have on any domain $\D$ with smooth boundary: 
\begin{eqnarray}\label{eq:wickrotation}
\int_{\D} \mathrm G^\alpha \TVolR \ &=&\lim_{\varepsilon \rightarrow 0^+}\frac{1}{\varepsilon}\int_{[t,t+\varepsilon]\times\D} \GBC^\alpha \nonumber \\
 &=& \lim_{\varepsilon \rightarrow 0^+}
\frac{1}{\varepsilon}\int_{\mathbf{u}([t,t+\varepsilon] \times \D)} \pi^*\GBC^\alpha \nonumber\\
&=&\lim_{\varepsilon \rightarrow 0^+} \frac{1}{\varepsilon}\left\{\left(\int_{\mathbf{u}(\D_{t+\varepsilon})}-\int_{\mathbf{u}(\D_{t})} \right) \boldsymbol \Pi^\alpha + \int_{\mathbf{u}([t,t+\varepsilon] \times \partial\D)} \boldsymbol\Pi^\alpha \right\}
\nonumber \\
&=&\lim_{\varepsilon \rightarrow 0^+} \frac{1}{\varepsilon}\left\{\left(\int_{\D_{t+\varepsilon}}-\int_{\D_{t}} \right) \mathbf{u}^*\boldsymbol \Pi^\alpha +\int_{[t,t+\varepsilon]\times \partial\D} \mathbf{u}^* \boldsymbol \Pi^\alpha \right\}
\nonumber \\&=& \int_{\D}\mathcal{L}_\mathbf{u}\left(\Pi^\alpha_{//}\boldsymbol\eta\right)+\int_{\partial\D} \Pi^\alpha_\perp \mathfrak{w} \nonumber
\\ \int_{\D} \mathrm G^\alpha \TVolR \  &=& \int_{\D} \left((\partial_t+\inI)\Pi_{//}^\alpha \right) \boldsymbol\eta\ + \int_{\partial\D} \Pi^\alpha_\perp \mathfrak{w} \ ,
\end{eqnarray}
where $\mathcal{L}_\mathbf{u}$ denotes the Lie derivative in the direction of $\mathbf u$.

\noindent
Beware: in order to get the third line in the computation above, one has to apply Stokes' theorem, but the boundary is not smooth. Therefore, one should first construct a sequence of  submanifolds with smooth boundary approximating $[t,t+\varepsilon]\times\D$ in `nice' fashion and then go to the limit. This works without creating an additional boundary term, 
because we chose $\bm\ell = \bm u$, fixed before the approximation is done.
One could  also try to apply the GBC formula by choosing a vector field $\bm\ell = \bm n$ normal to the boundary of the domain $[t,t+\varepsilon]\times \D$; this yields an additional term given by the `angle' of the boundary on the pieces $\{t,t+\varepsilon\}\times \partial \D$. We expect this other integration method to provide independent formulae in which the Euler characteristic of the boundary $\partial \D$ would play a role.

Recalling Avez' lemma (\ref{equ:avez}), and  
\begin{equation}
\label{TvolL}
\TVolL^\alpha=|\alpha|^{1/2}\TVolL \ ,
\end{equation} 
we obtain:
\begin{equation}
     \forall \alpha\in \R^*,\quad   \GBC^\alpha= |\alpha|^{3/2}\alpha^{-14} \CalA(\alpha)\TVolL \ .
\end{equation}
The general argument by Avez is that, up to some power of $|\alpha|$, the Gauss-Bonnet-Chern integrand is a 
rational\footnote{By rational in $\alpha$ we mean that the function admits an infinite sum expansion in $\alpha$ with coefficients in $C^{\infty} (\M,\mathbb{R}$) of the form $\sum_{n=p}^{+\infty}f_n \alpha^n$ with $p\leq 0$, and each $f_n$ is a scalar field on $\M$.} (hence meromorphic) function in the variable $\alpha$ with real coefficients depending on $\CmetricL_{ij}, \partial_k\CmetricL_{ij}$ and $\partial_{kl}\CmetricL_{ij}$, whereas the Euler characteristic is a constant.
We note that, for $f(\alpha)$ a rational function defined on $\R$ but at some finitely many (putative)  poles, if $f(\alpha)$ has constant value, {$b\in \R$,} on some open domain, say $\U$, then $f(\alpha)=b$ for all $\alpha\in \R$. The form $\GBC^\alpha$ has rational coefficients up to some power of $|\alpha|$, and we can generalize the previous argumentation to such functions. 

We wish to apply the previous line of argumentation to the following expressions.
\begin{eqnarray}
f(\alpha) &:=& \int_{\D} \mathrm{G}^\alpha \TVolR \ ;
\\g(\alpha) &:=& \int_{\D} \left((\partial_t+\inI)\Pi_{//}^\alpha \right) \boldsymbol\eta\ + \int_{\partial\D} \Pi^\alpha_\perp \mathfrak{w} \ .
\end{eqnarray}  
Following from Avez' lemma, we can decompose $f(\alpha)$ into a product of a power of $|\alpha|$ and a rational function; we give a more physical decomposition below (see lemma \ref{lem:interior}).
To obtain the similar assertion of $g(\alpha)$, we will follow the same line of argumentation as given by Avez:
\begin{itemize}
\item as shown by Avez, the Riemann tensor is a rational function in $\alpha$;
\item we recall \eref{TvolL};
\item using a chart described in the introduction of this subsection we get:
\begin{equation}
\forall i,j \in \{1,2,3\},\quad  (\nabla^{\alpha})_{j}(u^\alpha)^i = |\alpha|^{-1/2}\nabla_j u^i \ .
\end{equation}
\end{itemize}
Finally, $\boldsymbol\phi_d$ involves three times $\TVolL$, one time $\mathbf{u}$ and $2d$
times $\nabla \mathbf{u}$, hence, the coefficients of  $|\alpha|^{3/2-d}\boldsymbol{\phi}_d^\alpha$ are rational functions in $\alpha$.
We will then decompose $\Pi_{//}$ and $\Pi_\perp$ with respect to $\boldsymbol{\phi}_0$ and $\boldsymbol{\phi}_1$ and express $\GBC$, $\boldsymbol{\phi}_0$ and $\boldsymbol{\phi}_1$ in terms of physical quantities to conclude our computation (see lemmata \ref{lem:princ} and \ref{lem:PIperp}).

\subsection{Description of the GBC form} 
\label{section:final}

	A description of the interior term $\GBC$ in terms of physical quantities is needed to obtain a meaningful formula.  
	We recall the definition of the Nomizu-Kulkarni product $\owedge$ of two symmetric $(0,2)-$tensors $\bold A$ and $\bold B$, written in components \cite{MR867684},
		\begin{equation}
		(\mathbf A\owedge \mathbf B)_{ijkl}=A_{ik}B_{jl}+B_{ik}A_{jl}-A_{il}B_{jk}-A_{jk}B_{il}\ , 
		\end{equation} 
		and the Ricci decomposition we will use in the following.
	\begin{theo}[Ricci Decomposition \cite{MR867684}]\label{theo:decomp} On an $(N+1)$-dimensional semi-Riemann\-ian manifold $(\M,\TmetricL)$, the following decomposition is orthogonal for the scalar product $A\cdot B := A_{ijk\cdots}B^{ijk\cdots}$:
		\begin{equation} 
		\TRiemL=\TWeyl+ {\mathbf L}+ \mathbf{S}   \ ,
		\end{equation}
		where $\TWeyl$ is the Weyl tensor of $\TmetricL$, and ${\mathbf L}+ \mathbf{S}$ the non-conformal part of $\TRiemL$, with
		\footnote{${\mathbf{L}} + \mathbf{S} = \TmetricL \owedge \mathbf{P}$, where $\mathbf{P}$ is the Schouten tensor.
		We may write these tensors in components: 
\begin{eqnarray}.
\fl  
		{{L}_{ijkl}}&=&\frac{\RicciScalL}{N(N+1)}\left(\CmetricL_{ik}\CmetricL_{jl}-\CmetricL_{il}\CmetricL_{jk}\right) ; \\
\fl
\CSchouten_{ijkl}&=&\frac{2\RicciScalL}{(N+1)(N-1)}\left(\CmetricL_{ik}\CmetricL_{jl}-\CmetricL_{il}\CmetricL_{jk}\right) + \frac{1}{N-1}\left(\CmetricL_{ik}\CRicL_{jl}+\CmetricL_{jl}\CRicL_{ik} - \CmetricL_{il}\CRicL_{jk}-\CmetricL_{jk}\CRicL_{il}\right) .
		\end{eqnarray}}
		\begin{eqnarray}		
				{{\mathbf L}}&=&\frac{\RicciScalL}{2N(N+1)}\TmetricL\owedge \TmetricL \ ; \\
		\TSchouten&=&\frac{1}{N-1}\TmetricL \owedge \left(\TRicL-\frac{\RicciScalL}{N+1}\TmetricL\right)\ .
		\end{eqnarray}				
\end{theo}
From the definition of $\GBC$ and the symmetries of the Riemann tensor, one obtains Lanczos'  formula \cite{lanczos}: 
\begin{equation}
	\GBC=\frac{-1}{32\pi^2}\left(|\TRiemL|^2-4|\TRicL|^2+\RicciScalL^2 \right)\TVolL \ ,
\end{equation}
		where $|\mathbf A|^2 :=A_{ijk\cdots}A^{ijk\cdots}$.
		Combined with the Ricci decomposition, theorem \ref{theo:decomp}, we then obtain:
		\begin{equation}
		\label{eq:interior1}
			\GBC=\frac{-1}{32\pi^2}\left(|\TWeyl|^2
			-4|\TRicL|^2+{\RicciScalL}^2
			+| {{\mathbf L}} |^2 +|\TSchouten|^2 \right)\TVolL\ .
		\end{equation}
	Using the Einstein equation (\ref{eq:einstein}), we can evaluate all terms in formula (\ref{eq:interior1}) 
	except for the Weyl tensor. The simplest two are (recall that $N=3$): 
\begin{eqnarray}
 \RicciScalL^2 &=& (\varrho + 4 \Lambda)^2 \ ; \\
 |\TRicL|^2 &=& \varrho^2+2\Lambda\varrho+4\Lambda^2  \ .
\end{eqnarray}		
		 To ease the computation for the last two, one can notice that for any symmetric $2-$tensors $\mathbf{A},\mathbf{B}$:
		\begin{eqnarray}
		\fl\qquad
			\left| \mathbf A\owedge \mathbf B\right| ^2 &=&  \left(A\owedge B \right)_{ijkl} \left(A\owedge B \right)^{ijkl} \nonumber \\
			&=& 4\left(A_{ij}A^{ij}B_{kl}B^{kl} + (A_{ij}B^{ij})^2 - A_{i}^{~j}A_{j}^{~k}B_{k}^{~l}B_{l}^{~i}
			 -A_{i}^{~j}B_{j}^{~k}A_{k}^{~l}B_{l}^{~i}\right) \ .
		\end{eqnarray}
		When $\mathbf A=   \TmetricL$, this reduces to 
		\begin{eqnarray}
			\left| \TmetricL\owedge \mathbf B\right| ^2  &=& 8 B_{kl}B^{kl} + 4(B_{i}^{~i})^2 \ .
		\end{eqnarray}
In particular this gives: 
	\begin{eqnarray}
		\left| {{\mathbf L}} \right | ^2 &=& \frac{1}{(2\cdot 3 \cdot 4)^2}\left(8 \CmetricL_{ij}\CmetricL^{ij}+4(\CmetricL_i^{~i})^2 \right) \RicciScalL^2
		= \frac{1}{6}(\varrho+4 \Lambda)^2  \ ;  \\
		|\TSchouten|^2&=&\frac{1}{2^2}  \left\{8\left(\CRicL_{ij}-\frac{\RicciScalL}{4}\CmetricL_{ij}\right)\left(\CRicL^{ij}-\frac{\RicciScalL}{4}\CmetricL^{ij}\right)+4\left(\RicciScalL-\frac{\RicciScalL}{4}\CmetricL_{i}^{~i}\right)\right\} 		\nonumber \\ 
&=& 2\left(\CRicL _{ij} \CRicL^{ij} - \frac{2\RicciScalL}{4}\CRicL_{i}^{~ i}+\frac{ \RicciScalL ^2}{4^2}\CmetricL_{i}^{~i}\right) 
= \frac{3}{2}\varrho^2 \ . 
	\end{eqnarray}
We thus obtain: 

\begin{lem} \label{lem:interior} There exists a function $\mathcal B$, rational in $\alpha$, with coefficients in $\mathcal C^{\infty}(\M, \mathbb{R})$ such that:
\begin{equation}\label{eq:interior}
		\GBC^\alpha=\frac{|\alpha|^{3/2}}{\pi^2}\mathcal B(\alpha) \boldsymbol \varpi \quad \quad  \forall \alpha\in \R^*
\end{equation}
			and
			\begin{equation}
			\mathcal B(-1) = \frac{-1}{32}\left(|\TWeyl|^2-\frac{4}{3}\varrho^2+\frac{4}{3}\Lambda\varrho+\frac{8}{3}\Lambda^2\right)\ .
		         \end{equation}
\end{lem}

	\subsection{Computation of the Chern's form components}

We would like to express $\Pi^\alpha_{//}$ in terms of physical quantities.\footnote{A computation of the Wick rotation of $\boldsymbol\Pi$ can be found in \cite{alty}. However, we wish to have a formula in terms of scalar invariants of the extrinsic curvature (here minus the expansion tensor $\Theta_{ij}$) that the aforementioned paper does not provide. Furthermore, our computation includes cases where the vector field $\bm u$ is not normal to the boundary.} Since $u_{i;j} =\Theta_{ij}$ it is expected than at least part of $\Pi^{\alpha}_{//}$ can be written in terms of scalar invariants of the expansion tensor. We shall see that it is indeed the case.

\begin{lem}\label{lem:princ} There exist functions $\mathcal C$ and $\mathcal D$, rational in $\alpha$, with coefficients in $\mathcal C^{\infty}(\M, \mathbb{R})$ such that:
\begin{equation}\label{eq:boundary1}
			\Pi_{//}^\alpha  = \frac{|\alpha|^{-1/2}}{\pi^2}\mathcal C(\alpha) + \frac{|\alpha|^{1/2}}{\pi^2} \mathcal D(\alpha) \ ,\nonumber \\
\end{equation}
and 
		\begin{eqnarray}\label{eq:boundary12}
			\mathcal C(-1)  &=& \frac{1}{2}\inIII \ ; \\ 
			\mathcal D(-1) &=&\frac{3}{4}\inIII-\frac{1}{4}
   (\partial_t +\inI) \cdot \inII+\Lambda \inI \ .
\end{eqnarray}
		\end{lem}
\textit{Proof:} We  first compute $\mathbf u^*\boldsymbol{\phi}_0$ (denoting $(u^\alpha)_{i;j}:=(\nabla^\alpha)_j(u^\alpha)_i$ and $u_{i;j}:=(u^{\alpha=-1})_{i;j}
			 =(\nabla^{\alpha=-1})_j(u^{\alpha=-1})_i$):
		\begin{eqnarray*}
		\fl	\quad\quad	
		  ((\mathbf{u}^\alpha)^*\boldsymbol\phi^\alpha_{0})_{c_1c_2c_3}&=&
		  (\CVolL^\alpha)_{a_1a_2a_3a_4}(\CVolL^{\alpha})^{b_1b_2b_3b_4}(u^\alpha)^{a_1}(u^\alpha)^{a_2}_{~;b_2}(u^\alpha)^{a_3}_{~;b_3} (u^\alpha)^{a_4}_{~;b_4}(\CVolL^\alpha)_{b_1c_1c_2c_3}\\
		  &=& 
		  |\alpha|^{3/2-4/2}\CVolR_{a_2a_3a_4}\CVolR^{b_2b_3b_4}u^{a_2}_{~;b_2}u^{a_3}_{~;b_3} u^{a_4}_{~;b_4}\CVolR_{c_1c_2c_3}\\
		  &=& |\alpha|^{-1/2}
		  \CVolR_{a_1a_2a_3}\CVolR^{b_2b_3b_4}\Theta^{a_1}_{~b_1}\Theta^{a_2}_{~b_2}\Theta^{a_3}_{~b_3}\CVolR_{c_1c_2c_3}\\
		(\mathbf{u}^\alpha)^*\boldsymbol\phi^\alpha_{0} &  =& 6 |\alpha|^{-1/2} \inIII \ \TVolR\ .
		\end{eqnarray*}
		To obtain the second line, recall that $u^{a_i}_{~;0}=0$ since the fluid is geodesic and, thus, non-zero terms in the sum are those with $b_1=0$ and the spacelike leaves being orthogonal to $\mathbf u$, $u^i\CVolL_{ic_1c_2c_3}=\CVolL_{0c_1c_2c_3}=\CVolR_{c_1c_2c_3}$.  We thus define $\mathcal C(\alpha):=\frac{1}{2}  \inIII$, $C(\alpha)$ independent of $\alpha$ here.
		
Secondly, we compute $u^*\boldsymbol{\phi^\alpha}_{1|T\Sigma_t}$ :\footnote{Beware that on the first three lines the summation is over $0,1,2,3$, while on the last line the summation as well as the antisymmetrization is reduced to $1,2,3$. Antisymmetrization is applied before taking the trace.}
	\begin{eqnarray}
		  \mathbf{u}^*\boldsymbol\phi^\alpha_{1|T{\Sigma_t}} &=& = \frac{1}{12}(u^*\boldsymbol{\phi^\alpha}_{1})_{ijk}\CVolR^{ijk}\TVolR    \nonumber\\&=& \frac{1}{6} 
		(\CVolL^\alpha)_{a_1a_2a_3a_4}(\CVolL^\alpha)^{lb_2b_3b_4}(u^\alpha)^{a_1}(u^\alpha)^{a_2}_{~;b_2} (\CRiemL^\alpha)^{a_3a_4}_{~~b_3b_4} (\CVolL^\alpha)_{lijk}\CVolR^{ijk} \TVolR \nonumber \\
		  &=& \frac{1}{2}|\alpha|^{3/2-2/2}
		\CVolR_{a_2a_3a_4}\CVolR^{b_2b_3b_4}\Theta^{a_2}_{~b_2} (\CRiemL^\alpha)^{a_3a_4}_{~~b_3b_4} \TVolR\  \nonumber 
		\\&=& -3 |\alpha|^{1/2}\, \Theta^{[i}_{~[i}(\CRiemL^\alpha)^{jk]}_{~~jk]} \TVolR \ . \nonumber
		\end{eqnarray}
We notice that $\TRiemL^{\alpha}$ is rational in $\alpha$, so that we can define $\mathcal D(\alpha):=\frac{3}{8} \Theta^{[i}_{~[i}(\CRiemL^\alpha)^{jk]}_{~~jk]} $. Then:
		\begin{eqnarray}
		\fl
\qquad \mathcal D(-1)&=& \frac{3}{8}\, \Theta^{[i}_{~[i}\CRiemL^{jk]}_{~~jk]} \nonumber\\
&=&\frac{1}{8}\left( \Theta _i^{~i}\CRiemL_{jk}^{~jk}+2 \Theta _{i}^{~j}\CRiemL_{jk}^{~ki}\right) \nonumber \\
\fl
		  &=&\frac{1}{8}( \Theta _i^{~i}\CRiemR_{jk}^{~~jk} + \Theta _i^{~i} \Theta _j^{~j} \Theta _k^{~k} -\Theta _i^{~i} \Theta _j^{~k} \Theta _k^{~j} \nonumber \\
		  &&\left. \  +2 \Theta _{i}^{~j}\CRiemR_{jk}^{~~ki} + 2 \Theta _i^{~j} \Theta _j^{~k} \Theta _k^{~i}-2 \Theta _i^{~j} \Theta _j^{~i} \Theta _k^{~k}\right) 
		  \nonumber \\
		  \fl
		  &=& \frac{1}{8}\left( \inI\RicciScalR-2 \Theta _{i}^{~j}\CRicR_{j}^{~i} +6\inIII\right)\ .
		 \end{eqnarray}
From equation \eref{eq:partialII} we have:
\begin{equation}
 \inI \RicciScalR-2 \Theta _i^{~j}\CRicR_j^{~i}=-2\left(\partial_t\inII+\inI \cdot\inII -\Lambda\inI\right) \ ,
\end{equation}
so that:
\begin{equation}
            \mathcal D(-1) = \frac{3}{4}\inIII-\frac{1}{4}\left( \partial_t \inII +\inI\cdot\inII-\Lambda \inI\right)\  .
          \end{equation}
$\Box$

\begin{lem} \label{lem:PIperp} There exists a function $\mathcal E$, rational in $\alpha$, with coefficients in $\mathcal{C}^\infty(\M, \mathbb{R})$ such that:
$$ \Pi_\perp^\alpha = \frac{|\alpha|^{1/2}}{\pi^2}\mathcal E(\alpha) \ ,$$ 
and $$\mathcal E(-1) = \frac{1}{4}H_{i}^{~j}w_j^{~k}\Theta_{k}^{~i} \ ,$$
with $H_{ij}$ the components of the spatially projected magnetic part of the Weyl tensor, and $w_{ij}$ the components of the volume form of the boundary $\partial\D$.
\end{lem}
\textit{Proof:}
From the definition of $\Pi_\perp$, and $( \underline{\boldsymbol{u}}  \wedge\mathfrak{w})_{abc}\CmetricL^{ai}\CmetricL^{bj}\CmetricL^{ck} ( \underline{\boldsymbol{u}}\wedge\mathfrak{w})_{ijk} = -6$ we have: 
\begin{eqnarray*}
\fl\qquad\qquad \Pi_\perp^\alpha &=& \frac{1}{6\cdot 8 \pi^2} \left((\mathbf{u}^\alpha)^*\boldsymbol \phi^\alpha_{1|T(\partial\D\times \R)}\right)_{abc}\CmetricL^{ai}\CmetricL^{bj}\CmetricL^{ck} (\underline{\boldsymbol{u} } \wedge\mathfrak{w})_{ijk}\\
\fl\qquad&=& 
\frac{1}{6\cdot 16 \pi^2}u^\alpha_{a_1}(\CVolL^\alpha)^{a_1a_2a_3a_4}\CRiemL^\alpha_{a_3a_4b_3b_4}(\CVolL^\alpha)^{b_1b_2b_3b_4}u^\alpha_{a_2;b_2}(\CVolL^\alpha)_{b_1ijk}u^iw^{jk}   
\\
\fl &=& -
\frac{|\alpha|^{1/2}}{16\pi ^2}u_{a_1}\CVolL^{a_1a_2a_3a_4}\CRiemL^\alpha_{a_3a_4b_3b_4}\CVolL^{b_1b_2b_3b_4} u_{a_2;b_2} v_{b_1}
\\
&=& 
\frac{|\alpha|^{1/2}}{16\pi ^2}u_{a_1}\CVolL^{a_1a_2a_3a_4}\CRiemL^\alpha_{a_3a_4b_3b_4}(u^{b_2}w^{b_3b_4}+u^{b_3}w^{b_4b_2}+u^{b_4}w^{b_2b_3}) u_{a_2;b_2} 
\\
&=& 
\frac{|\alpha|^{1/2}}{8\pi ^2}\CVolR^{a_2a_3a_4}\CRiemL^\alpha_{a_3a_4b_3b_4}u^{b_3}w^{b_4b_2} \Theta_{b_2a_2} 
\\
\fl &=& 
\frac{|\alpha|^{1/2}}{\pi^2} \mathcal{E}(\alpha)\ .
\end{eqnarray*}
With $\mathcal E(\alpha)=
\frac{1}{8}\CVolR^{a_2a_3a_4}\CRiemL^\alpha_{a_3a_4b_3b_4}u^{b_3}w^{b_4b_2} \Theta_{b_2a_2} $,
which is a rational function in $\alpha$, we have: 
$$\mathcal E(-1)=
\frac{1}{8}\CVolR^{a_2a_3a_4}\CRiemL_{a_3a_4b_3b_4}u^{b_3}w^{b_4b_2} \Theta_{b_2a_2}\ .$$
We now use the Ricci decomposition to obtain:
$\TRiemL = \TWeyl + \TmetricL \owedge \mathbf{P}$,
where $\mathbf{P}$ is the Schouten tensor.
Using the definition of the Nomizu-Kulkarni product, we obtain:
\begin{eqnarray*}
\fl\qquad
\CVolR^{a_2a_3a_4}(\CmetricL\owedge P)_{a_3a_4b_3b_4}u^{b_3}w^{b_4b_2} \Theta_{b_2a_2} \\ \fl \qquad\qquad = \CVolR^{a_2a_3a_4}\CmetricL_{a_3b_3} P_{a_4b_4}w^{b_4b_2} u^{b_3}\Theta_{a_2b_2}
+\CVolR^{a_2a_3a_4}\CmetricL_{a_4b_4}P_{a_3b_3}w^{b_4b_2} u^{b_3}\Theta_{a_2b_2}
\\ \fl \qquad \qquad\quad
-\ \CVolR^{a_2a_3a_4}\CmetricL_{a_3b_4} P_{a_4b_3}w^{b_4b_2} u^{b_3}\Theta_{a_2b_2}
-\CVolR^{a_2a_3a_4}\CmetricL_{a_4b_3}P_{a_3b_4}w^{b_4b_2} u^{b_3}\Theta_{a_2b_2} .
\end{eqnarray*}
From Einstein's equation we have:
$$ P_{ij} = A\CmetricL_{ij}+Bu_iu_j \ , $$
where $A$ and $B$ are some scalar fields, and
since $\CVolR^{ijk}u_k = 0$, all four terms above vanish.
We thus have:
$$\CVolR^{a_2a_3a_4}(\CmetricL\owedge P)_{a_3a_4b_3b_4}w^{b_4b_2} u^{b_3}\Theta_{a_2b_2} =0 \ ,$$
and finally:
\begin{eqnarray*}
\fl\qquad
\mathcal E(-1) =
\frac{1}{8}\CVolR^{a_2a_3a_4}\CWeyl_{a_3a_4b_3b_4} u^{b_3}w^{b_4b_2}\Theta_{b_2}^{~a_2}
=\frac{1}{4}H_{a_2b_4}w^{b_4b_2}\Theta_{b_2a_2}=\frac{1}{4}H_{i}^{~j}w_{j}^{~k}\Theta_{k}^{~i} \ .
\end{eqnarray*} 
$\Box$

\section{Results and Discussion}
\label{section:Results}

We summarize the results of the previous section in terms of the main theorem of this article, and develop a number of corollaries to this theorem, verify the results in terms of derivations from a class of exact solutions, and discuss the resulting evolution equations for the backreaction variables.

\subsection{Final formula}

Combining lemmata \ref{lem:interior}, \ref{lem:princ}, \ref{lem:PIperp} and equation \eref{eq:wickrotation}, we obtain for all $\alpha>0$: 
\begin{eqnarray}
\fl
 \int_{\D} \mathrm G^\alpha \TVolR \  &=& \int_{\D} \left((\partial_t+\inI)\Pi_{//}^\alpha \right) \boldsymbol\eta\ + \int_{\partial\D} \Pi^\alpha_\perp \mathfrak{w}
\nonumber \ ;\\ \nonumber 
\fl \frac{|\alpha|^{3/2}}{\pi^2} \langle\mathcal B (\alpha)\rangle_\D &=& \frac{|\alpha|^{-1/2}}{\pi^2}(\partial_t+\inI)\langle\mathcal C(\alpha)\rangle_\D+\frac{|\alpha|^{1/2}}{\pi^2}(\partial_t+\inI)\langle\mathcal D(\alpha)\rangle_\D + \frac{|\alpha|^{1/2}}{\pi^2}\frac{|\partial \D|}{|\D|} \langle \mathcal E(\alpha) \rangle_{\partial\D} \ ; \nonumber\\
\fl
\alpha^2 \langle\mathcal B (\alpha)\rangle_\D & =& (\partial_t+\inI)\langle\mathcal C(\alpha)\rangle_\D+\alpha (\partial_t+\inI)\langle\mathcal D(\alpha)\rangle_\D + \alpha\frac{|\partial \D|}{|\D|} \langle \mathcal E(\alpha) \rangle_{{\partial\D}} \quad \left(\forall\alpha>0\right) \ .\nonumber
\end{eqnarray}
Since $\mathcal B(\alpha), \mathcal C(\alpha), \mathcal D(\alpha),\mathcal E(\alpha) $ are all rational in $\alpha$, so are their averages, and since the last equation is an equality between rational functions, it is therefore valid for every $\alpha\in \R$ which is not a pole. We finally have: 
\begin{equation}
\fl\qquad
 \langle\mathcal B (-1)\rangle_\D  = (\partial_t+\inI)\langle\mathcal C(-1)\rangle_\D- (\partial_t+\inI)\langle\mathcal D(-1)\rangle_\D - \frac{|\partial \D|}{|\D|} \langle \mathcal E(-1) \rangle_{{\partial\D}} \ .
\end{equation}

\begin{theo}\label{theo:final}
	Let $\D$ be a compact spacelike hypersurface with smooth boundary in a $4-$dimensional Lorentzian manifold $\M$ satisfying the irrotational dust Einstein equation. If $\D$ is fluid-orthogonal, then the following holds:\footnote{Recall that for any scalar $\psi$: $$(\partial_t + \langle\inI\rangle)^2 \psi:= (\partial_t +\langle\inI\rangle) (\partial_t+\langle\inI\rangle) \psi = \partial_{tt}\psi + \langle\inI\rangle \partial_t \psi + \partial_t(\langle\inI\rangle\psi) +  \langle\inI\rangle^2 \psi = \left(\partial_{tt} + 2\langle\inI\rangle\partial_t + \partial_t\langle\inI\rangle +  \langle\inI\rangle^2\right)\psi \; .$$}
		\begin{eqnarray}\label{mainformula}
		\fl\ \
			\frac{1}{8}\langle |\TWeyl|^2\rangle_\D-\frac{1}{6}\langle\varrho^2\rangle_{\D}+\frac{1}{6}\Lambda\langle\varrho\rangle_\D+\frac{1}{3}\Lambda^2 \nonumber\\
	\fl\ \		=\left(\partial_t+\langle\inI\rangle_\D\right)\langle \inIII \rangle_{\D}-(\partial_t+\langle \inI\rangle_\D)^2 \langle\inII\rangle_\D 
		 +\Lambda(\partial_t+\langle \inI \rangle_\D)\langle \inI\rangle_\D + \frac{|\partial \D|}{|\D|}\langle H_{ij}w^{jk}\Theta_{k}^{~i} \rangle_{\partial\D}\, .
		\end{eqnarray}
\end{theo}

\begin{cor}\label{cor1}
	Under the same conditions as in theorem \ref{theo:final}, and if $\M$ is `silent', i.e. $H_{ij}=0$, we have: 
	\begin{eqnarray}
	\frac{1}{8}\langle |\TWeyl|^2\rangle_\D-\frac{1}{6}\langle\varrho^2\rangle_\D+\frac{1}{6}\Lambda\langle\varrho\rangle_\D+\frac{1}{3}\Lambda^2 \nonumber \\ =\left(\partial_t+\langle\inI\rangle_\D\right)\langle \inIII \rangle_\D-(\partial_t+\langle \inI\rangle_\D)^2 \langle\inII\rangle_\D +\Lambda(\partial_t+\langle \inI \rangle_\D)\langle \inI\rangle_\D\ .
	\end{eqnarray}
\end{cor}

\begin{cor}\label{cor2}
	Let $\Sigma$ be a compact and without boundary spacelike hypersurface of a $4-$dimensional Lorentzian manifold $\M$ satisfying the irrotational dust Einstein equation. If $\Sigma$ is fluid-orthogonal, then the following holds:
	\begin{eqnarray}
	\frac{1}{8}\langle |\TWeyl|^2\rangle_\Sigma-\frac{1}{6}\langle\varrho^2\rangle_{\Sigma}+\frac{1}{6}\Lambda\langle\varrho\rangle_\Sigma+\frac{1}{3}\Lambda^2 \nonumber \\ =\left(\partial_t+\langle\inI\rangle_\Sigma\right)\langle \inIII \rangle_{\Sigma}-(\partial_t+\langle \inI\rangle_\Sigma)^2 \langle\inII\rangle_\Sigma +\Lambda(\partial_t+\langle \inI \rangle_\Sigma)\langle \inI\rangle_\Sigma\ .
	\end{eqnarray}
\end{cor}			

\noindent
We proceed with a remark on the boundary term, absent in the last two corollaries. As is expected, the boundary term is a full divergence.

\begin{lem} The boundary integrand is a full covariant divergence, more precisely: 
for any domain $\D$ with  smooth boundary we have:
\begin{equation}
\int_{\D}(H_{ij}\Theta_k^{~i}\eta^{jkl})_{|| l}\; \TVolR = \int_{\partial\D} H_{ij}\Theta_k^{~i}w^{jk}\; \mathfrak{w}\ .
\end{equation}
\end{lem}

\noindent
The main theorem thus has a corresponding local formulation.
\begin{cor}\label{cor:cor3}
Under the same hypothesis as theorem \ref{theo:final} one has:
\begin{eqnarray}
		\fl
			\frac{1}{8}|\TWeyl|^2-\frac{1}{6}\varrho^2+\frac{1}{6}\Lambda\varrho+\frac{1}{3}\Lambda^2 &=&\left(\partial_t+\inI\right) \inIII -(\partial_t+ \inI)^2 \inII  +\Lambda(\partial_t+ \inI ) \inI +(H_{ij}\Theta_k^{~i}\eta^{jkl})_{|| l}\ .
			\nonumber \\
		\end{eqnarray}
\end{cor}

\noindent
In what follows we shall abbreviate the average of the boundary term,
\begin{equation}
\mathfrak{B}_\D : =  \frac{|\partial \D|}{|\D|}\langle H_{ij}w^{jk}\Theta_{k}^{\ i} \rangle_{\partial\D} =
\langle (H_{ij}\Theta_k^{~i}\eta^{jkl})_{|| l}\rangle_\D \ .
\end{equation}
\vspace{10pt}
\noindent
We are now going to investigate alternative forms of the theorem that feature the relevant variables in the cosmological context. Using the averaged energy constraint \eref{hamiltonav}, and verifying the equalities
$(\partial_t+\langle\inI\rangle_\D)^2 \langle\varrho\rangle_\D= 0$ and $(\partial_t+\langle\inI\rangle_\D)^2 \Lambda= \Lambda(\partial_t+\langle\inI\rangle_\D) \langle\inI\rangle_\D$, we obtain the following compact relation for the evolution of the averaged scalar curvature.

\begin{cor} \label{cor4}
\textit{Under the same conditions as in theorem \ref{theo:final}, the following second-order evolution equation for the averaged scalar curvature holds:}
\begin{equation}
\fl
\frac{1}{2}(\partial_t+\langle \inI\rangle_\D)^2 \langle\RicciScalR\rangle_\D + \left(\partial_t+\langle\inI\rangle_\D\right)\langle \inIII \rangle_{\D}
= \frac{1}{8}\langle |\TWeyl|^2\rangle_\D -\frac{1}{6}\langle\varrho^2\rangle_{\D}+\frac{1}{6}\langle\varrho\rangle_\D \Lambda +\frac{1}{3}\Lambda^2 - \mathfrak{B}_\D  .
\end{equation}
\textit{In expanded form (using the Hubble functional $H_\D = 1/3 \average{\inI}$):}
\begin{eqnarray}
\fl\qquad
\ddotaverage{\CR}+ 6  H_\D \dotaverage{\CR}+ (3 \dot{H}_\D + 9  H_\D^2)\average{\CR} + 
2\left(\partial_t+3 H_\D\right)\langle \inIII \rangle_{\D} \nonumber\\
= \frac{1}{4}\langle |\TWeyl|^2\rangle_\D -\frac{1}{3}\langle\varrho^2\rangle_{\D}+\frac{1}{3}\langle\varrho\rangle_\D \Lambda +\frac{2}{3}\Lambda^2 -2 \mathfrak{B}_\D  \ .
\label{Revolution}
\end{eqnarray}
\end{cor}

We can further simplify the expanded form, equation \eref{Revolution}, by inserting a combination of the averaged 
energy constraint \eref{equ:ham} and the averaged Raychaudhuri equation \eref{equ:raych} (for N=3).\footnote{\label{combi}Useful combinations of the averaged energy constraint,  $3H_\D^2 = \average{\varrho} + \Lambda - \frac{1}{2} (\CQ_\D + \average{\CR})$, and the averaged Raychaudhuri equation, $3 \dot{H}_\D = -\frac{3}{2}\average{\varrho} + \frac{3}{2}\CQ_\D + \frac{1}{2} \average{\CR}$, are the following: $3({\dot H}_\D + H_\D^2) = -\frac{1}{2}\average{\varrho} + \Lambda + \CQ_\D$ (independent of $\average{\CR})$, and ${\dot H}_\D + 3 H_\D^2 = \frac{1}{2}\average{\varrho} + \Lambda - \frac{1}{3} \average{\CR}$ (independent of $\CQ_\D$).}
We also enforce the variance, $\average{\varrho^2} - \average{\varrho}^2$, in favour of just $\average{\varrho^2}$ and with
\begin{eqnarray}
\fl
-\frac{1}{3} \average{\varrho^2} +\frac{1}{3} \average{\varrho}\Lambda + \frac{2}{3}\Lambda^2 
= -\frac{1}{3}\left[ \average{\varrho^2}\; -\; \average{\varrho}^2 \right] -\frac{2}{3} \left[\average{\textstyle\frac{\varrho}{2}}-\Lambda\right] \left[\average{\varrho}+ \Lambda\right]  \, ,
\end{eqnarray}
we obtain the following alternative form.

\begin{cor} \label{cor5}
\textit{Under the same conditions as in theorem \ref{theo:final}, the following second-order evolution equation for the averaged scalar curvature holds:}
\begin{eqnarray}
\fl\qquad
\ddotaverage{\CR}+ 6  H_\D \dotaverage{\CR}+ \left(3 \left[\average{\textstyle\frac{\varrho}{2}} + \Lambda \right] -\average{\CR} \right) \average{\CR}
+ \frac{2}{3} \left[\average{\textstyle{\frac{\varrho}{2}}}-\Lambda\right] \left[\average{\varrho}+ \Lambda\right] \nonumber\\
\fl\quad
= -  2\left(\partial_t+3 H_\D\right)\langle \inIII \rangle_{\D} + \frac{1}{4}\langle |\TWeyl|^2\rangle_\D -\frac{1}{3}\left[\langle\varrho^2\rangle_{\D} - \average{\varrho}^2\right]
 -2 \mathfrak{B}_\D  \; .
\label{Revolution2}
\end{eqnarray}
\end{cor}

In the form of corollary \ref{cor4} we can provide the evolution equation for the kinematical backreaction $\CQ_\D$
by exploiting the integrability condition \eref{equ:integrab} (for $N=3$).
We transform the time-derivative operator $\ddotaverage{\CR}+6H_\D\dotaverage{\CR}$ by making use of the integrability condition and its time-derivative (using ${\dot a}_\D / a_\D =H_\D$):
\begin{eqnarray}
\fl\qquad
\label{eq:intconst}0&=&\dotaverage{\CR}+2H_\D\average{\CR}+\dot{\CQ}_\D+6H_\D\CQ_\D =: {\mathcal I} \\
\fl\qquad
0&=&\ddotaverage{\CR}+2H_\D\dotaverage{\CR}+2\dot{H}_\D\average{\CR}+\ddot{\CQ}_\D+6H_\D\dot{\CQ}_\D+6\dot{H}_\D\CQ_\D = \dot{\mathcal I} \ ,
\end{eqnarray}
i.e., we can construct various combinations of conservation laws, the relevant here being $\dot{\mathcal I} + 4H_\D \, {\mathcal I}=0$, which translates into the following equality: 
\begin{equation}
\fl
\ddotaverage{\CR}+6H_\D\dotaverage{\CR}+ (2 \dot H_\D + 8 H_\D^2) \average{\CR} = -\ddot{\CQ}_\D -10H_\D\dot{\CQ}_\D - (6 \dot H_\D + 24 H_\CD^2) \CQ_\D \, .
\end{equation}
Using the operator $(\partial_t+3H_\D)$ we obtain the equivalent equation:
\begin{eqnarray}
\fl
(\partial_t+3H_\D)^2\average{\CR} - (\dot H_D +H_\D^2) \average{\CR} = \nonumber\\
 - (\partial_t+3H_\D)^2\CQ_\D -4 H_\D (\partial_t+3H_\D)\CQ_\D - 3 (\dot H_\D + H_\D^2) \CQ_\D  \ .
\end{eqnarray}
Computing the last expression by using again combinations of the averaged Raychaudhuri equation \eref{equ:raych} and the averaged energy constraint \eref{equ:ham} (for $N=3$; \textit{cf.} footnote~\ref{combi}), we obtain the following from corollary~\ref{cor4} ($H_\D = 1/3 \average{\inI}$). 

\begin{cor}\label{cor6}
\textit{Under the same conditions as theorem \ref{theo:final}, the following second-order evolution equation for the kinematical backreaction variable $\CQ_\D =   \frac{2}{3}\langle \inI \rangle_\D^2 - 2 \langle\inII\rangle_\D$ holds:}
\begin{eqnarray}
\fl\quad\;
\left(\partial_t+\average{\inI}\right)^2\CQ_\D + \frac{4}{3}\average{\inI} \left(\partial_t+\average{\inI}\right)\CQ_\D + 
\frac{4}{3}\left(\frac{1}{6}\average{\inI}^2 + \frac{\Lambda}{2} - \average{\varrho} + \CQ_\D \right) \CQ_\D 
\nonumber \\
\fl
= 2(\partial_t+\average{\inI})\langle \inIII \rangle_{\D} - \frac{2}{9}\average{\inI}^2 \left[\Lambda - \average{\textstyle{\frac{\varrho}{2}}}\right]
- \frac{1}{4}\langle |\TWeyl|^2\rangle_\D +\frac{1}{3}\left[\langle\varrho^2\rangle_{\D}-\langle\varrho\rangle_{\D}^2\right] -2 \mathfrak{B}_\D \ .
\label{Qequation}
\end{eqnarray}
\textit{In expanded form (also following directly from the expanded form \eref{Revolution2}):}
\begin{eqnarray}
\fl\quad \ddot\mathcal{Q}_\D +10 H_\D \dot\mathcal{Q}_\D + \left( 20 H_\D^{2} -\frac{2}{3}\left[\average{\varrho} + \Lambda\right] + \frac{7}{3} \left[ \Lambda - \average{\textstyle{\frac{\varrho}{2}}} + \mathcal{Q}_\D \right] \right)\mathcal{Q}_\D   \nonumber \\ 
\fl
= 2(\partial_t+3H_\D)\langle \inIII \rangle_{\D} - 2  H_\D^{2}\left(\Lambda - \average{\textstyle{\frac{\varrho}{2}}}\right)
- \frac{1}{4}\langle |\TWeyl|^2\rangle_\D +\frac{1}{3}\left[\langle\varrho^2\rangle_{\D}-\langle\varrho\rangle_{\D}^2\right] -2 \mathfrak{B}_\D \ .
\label{Qexpanded}
\end{eqnarray}
\end{cor}

\vspace{10pt}
\noindent
Before we discuss the above theorem and its corollaries, we shall explicitly check special cases through derivation from a class of exact solutions.

\subsection{Silent model universes and the Szekeres family of solutions, a verification}

Silent model universes \cite{marco:silent,maeda:quiet,henk:silent,bolejko:silent} furnish an easy way to check the main theorem (more precisely corollary \ref{cor1})  for a wide class of solutions, since they satisfy a set of ordinary differential equations. These examples include FLRW solutions as well as spherically symmetric LTB (Lema\^\i tre-Tolman-Bondi) solutions \cite{krazinski,bolejko:review,sussman1,sussman2} and the more general (here pressure-free) Szekeres solutions \cite{szekeres1975class,krazinski,sussman3,PKbook} which satisfy all the hypotheses outlined in section \ref{section:hypothesis}.

We start by demonstrating the theorem \ref{theo:final} for the class of locally isotropic FLRW solutions. For this class the Weyl tensor vanishes (see standard textbooks, e.g. \cite{HT}) and the theorem reduces to the following expression:
$$-\frac{1}{6}\varrho_H^2 + \frac{1}{3}\Lambda^2 +\frac{1}{6}\Lambda \varrho_H=\left(\partial_t+\inI_H\right) \inIII_H-(\partial_t+ \inI_H)^2 \inII_H + \Lambda(\partial_t+ \inI_H)\inI_H \ ,$$ where we have put $\inI = \inI_H (t)= 3H = 3\frac{\dot a}{a}$ with the scale factor $a(t)$; $\inII =\inII_H (t)= 3H^2$; $\inIII = \inIII_H (t)= H^3$, and $\varrho = \varrho_H (t)$, obeying the continuity equation $\dot \varrho_H = -3H \varrho_H$ and the Friedmann equation $\varrho_H = 3H^2-\Lambda + 3k/a^2$, with the curvature constant $k$.
Using these relations, one directly gets: $$\dot H = -\frac{3}{2}H^2+\frac{1}{2}\Lambda - \frac{1}{2}\frac{k}{a^2}\ ; \ \ddot H = \frac{9}{2}H^3 -\frac{3}{2}\Lambda H +\frac{5}{2}\frac{k}{a^2} H\ .$$ Thus, expanding the right-hand side of the theorem's formula, one obtains:
\begin{eqnarray*}
\mathrm{RHS}&=& 
24 \, H^{4} + 9 \, \Lambda H^{2} - 42H^{2}\dot H - 3\Lambda \dot H - 6 \, \dot H^{2} - 6 \, H \ddot H
\\&=&-\frac{3}{2} \, H^{4} + \frac{3}{2} \, \Lambda H^{2} - 3 \, k H^{2} + \frac{3}{2} \, \Lambda k - \frac{3}{2} \, k^{2}
\\ &=& -\frac{1}{6}\varrho_H^2 + \frac{1}{3}\Lambda^2 +\frac{1}{6}\Lambda \varrho_H \ ,
\end{eqnarray*} 
which verifies the theorem. $\Box$

\medskip

We now move to the general system of silent model universes.
A specificity of this class of models and the corresponding classes of exact solutions is that the gravitoelectric part of the Weyl tensor $\mathbf E$ and  the shear tensor $\boldsymbol\sigma$ are
co-diagonalizable, and their eigenvalues together with $\varrho$ and $\inI$ satisfy a system of ordinary differential equations. Furthermore, in these cases, $\mathbf{E}$ and $\boldsymbol \sigma$
are scalars. We call $\mathcal E$  and $\s$ the respective eigenvalues of $\mathbf E$ and $\boldsymbol \sigma$ in the direction of the orbit of the $\mathrm{SO}(3)$-action.
One obtains the following system of ordinary differential equations (for their derivation see the references above):
 	\begin{eqnarray}	\label{silent}
	            \dot\varrho &=&-\varrho \inI \ ;   \nonumber \\		   
	            \dot\inI &=&-\frac{1}{3}\inI^2-\frac{1}{2}\varrho-6\s^2 +\Lambda \ ;\nonumber   \\
		    \dot \s &=&-\frac{2}{3}\inI \s+\s^2-\mathcal E \ ; \nonumber \\
		    \dot{\mathcal E} &=&-\frac{1}{2}\varrho\s - 3\mathcal E\s   - \mathcal E \inI \ . 
 	\end{eqnarray}
Since we have
\begin{equation}
\label{silentaux}
\inII=\frac{1}{3}\inI^2 - 3\s^2 \quad;\quad 
\inIII = \frac{1}{27}\inI^3 -  \inI \s^2 -2\s^3 \quad;\quad 
|\TWeyl|^2=48\mathcal{E}^2 \ ,
\end{equation}
a direct computation using equations \eref{silent} and \eref{silentaux} provides:
  \begin{eqnarray}
\label{silentdiff} 
    \dot\inII &=& -\frac{2}{9}\inI^3 - 6 \s^3 + 6 \mathcal{E}\s - \frac{1}{3}\inI\varrho + \frac{2}{3}\Lambda \inI\ ; \nonumber \\
    \dot\inIII &=& -\frac{1}{27}\inI^4 + 2 \inI\s^3 + 2 \inI \mathcal{E} \s
     + (\inI^2 + 6\mathcal{E})\s^2 - \frac{1}{18}(\inI^2 - 9\s^2)\varrho \nonumber \\  && \quad +\frac{1}{9}\Lambda \inI^2 - \Lambda\s^2  \ ; \nonumber \\
\ddot \inII &=&  \frac{2}{9}\inI^4 + 12\inI\s^3 - 18\s^4 - 10\inI\mathcal{E}\s + 2(2\inI^2 + 3\mathcal{E})\s^2 -6\mathcal{E}^2    
    \nonumber \\ 
    &&\quad+ \frac{1}{9}(7\inI^2 - 9\s^2)\varrho + \frac{1}{6}\varrho^2 - \frac{8}{9}\Lambda \inI^2-\frac{2}{3}\Lambda\varrho - 4\Lambda \s^2 + \frac{2}{3}\Lambda^2\ .
   \end{eqnarray}
Equations \eref{silent}, \eref{silentaux} and \eref{silentdiff} can be combined into a single  equation:
      \begin{eqnarray}
   \fl (\partial_t+\inI)\inIII-(\partial_t+\inI)^2 \inII+\Lambda(\partial_t+\inI) \inI -\frac{1}{8}|\TWeyl|^2+\frac{1}{6}\varrho^2\; -\frac{1}{6}\Lambda\varrho-\frac{1}{3}\Lambda^2= \;0 \ .
   \end{eqnarray}
Using the commutation rule in the form \eref{comm3}, we obtain theorem \ref{theo:final} for the Szekeres family of solutions:
   \begin{eqnarray}
   \fl (\partial_t+\langle\inI\rangle_\CD)\langle\inIII\rangle_\CD-(\partial_t+\langle\inI\rangle_\CD)^2 \langle\inII\rangle_\CD+\Lambda(\partial_t+\langle\inI\rangle_\CD)\langle \inI\rangle_\CD &&\nonumber\\\quad\quad\quad\quad-\frac{1}{8}\langle|\TWeyl|^2\rangle_\CD+\frac{1}{6}\langle\varrho^2\rangle_\CD\; -\frac{1}{6}\Lambda\langle\varrho\rangle_\CD-\frac{1}{3}\Lambda^2= \;0 \ .
   \end{eqnarray}
We conclude that the main theorem \ref{theo:final} holds for silent model universes containing the Szekeres family of solutions. 

\subsection{Discussion of theorem \ref{theo:final} and of its corollaries}

Theorem \ref{theo:final} and its corollaries supplement the non-closed system of scalar-averaged equations by one further equation for the volume scale factor $a_\D$, the averaged dust density $\average{\varrho}$, the averaged scalar curvature $\average{\CR}$, and the kinematical backreaction $\CQ_\D$. Although only one further equation is needed to close the system, the new equation furnished by corollary \ref{cor6} to theorem \ref{theo:final} for the kinematical backreaction variable $\CQ_\D$ contains further variables for which further evolution equations have to be derived, as is expected.
We also note that, though the boundary term has the form discussed in section \ref{sec:2Ddiscussion}, the formula obtained does not give a ``static" relation between $\average{\CR}$ and boundary terms but rather a ``dynamical'' relation. Indeed, we obtain a second-order differential equation relating $\average{\CR}$ to source terms and a boundary term in corollary \ref{cor5}.

We see a number of advantages of the particular form of these new equations, since this defines a promising route to a physical closure condition on this level of the hierarchy of scalar-averaged equations. 
To underpin this statement, we look at the equation \eref{Qexpanded}.
We have written this equation so that all terms in the first line contain known variables, $\CQ_\D$, $H_\D$ and $\average{\varrho}$, while the right-hand side of this equation is still to be evaluated. We identify two expressions: the first two terms in the second line feature a time-evolving expression for the averaged third principal scalar invariant of the expansion tensor (here divided by $2$):
\begin{equation}
\label{topconservation}
\fl\qquad \average{\inIII}^{\bdot} +3H_\D \average{\inIII} - H_\D^{2}\left[\Lambda - \average{\textstyle{\frac{\varrho}{2}}}\right]= \frac{1}{V_\D}\left({\average{\inIII}} V_\D \right)^{\bdot} - H_\D^{2}\left[\Lambda - \average{\textstyle{\frac{\varrho}{2}}}\right] \ ,
\end{equation}
with the volume $|\D| = V_\D$. This expression is zero for the homogeneous FLRW case. All other terms also vanish in the FLRW case, the non-local terms $\CQ_\D = 0$, $\langle\varrho^2\rangle_{\D}-\langle\varrho\rangle_{\D}^2 = 0$, and already before averaging, the local terms $|\TWeyl|^2 = 0$, $H_{ij} = 0$.
For the above expression there are indications in Newtonian theory that the averaged third invariant multiplied by the volume, $\average{\inIII} V_\D$, is related to the Euler characteristic of the domain boundary $\partial\D$, see [section~3.1.2]\cite{buchert:review}, a property that has yet to be understood within general relativity. This calls for a relation of a topological conservation law for the boundary of the averaging domain to the expression \eref{topconservation}.

The next terms on the second line on the right-hand side of this equation feature the averaged Weyl invariant, the variance of the dust density, and a boundary term; we here write the first two of these terms,
\begin{equation}
- \frac{1}{4}\average{|\TWeyl|^2} +\frac{1}{3}\left(\average{\varrho^2}-\average{\varrho}^2\right) \ .
\end{equation}
For this expression there are indications on a relation of the averaged Weyl invariant to information entropy functionals of the density and the kinematical backreaction, see \cite{Weylentropy}, a relation that is currently generalized by employing relativistic Lagrangian perturbation theory \cite{lagpenrose}.

Ongoing work is dedicated to understand and model these terms, before looking at further evolution equations down the hierarchy of scalar-averaged equations.
It will be nonetheless necessary to continue the search for new equations in the hope to close the scalar-averaged system.  In this regard, we followed an intricate method to obtain the main formula \eref{mainformula} of theorem~\ref{theo:final}. To the best of our knowledge this formula is new. However, considering corollary \ref{cor:cor3}, it should be possible to derive this formula directly from the general hierarchy of $3+1$ Einstein equations. 
We put complementary formulae into perspective in \ref{app:closure} that provide further insights. Looking at new ways to find such formulae we note that there are unexplored degrees of freedom in the method: the vector field $\boldsymbol{\ell}$ in Chern's formula in theorem \ref{theo:Chern} is free; we computed a formula identifying $\boldsymbol{\ell}$ with the fluid $4-$velocity which was the simplest choice. Other choices can be made to extract other meaningful information. Furthermore, the Gauss-Bonnet-Chern formula is but one example among many such integral formulae, and we expect that further equations could be derived from formulae coming from the second Chern's class or the Atiyah-Patodi-Singer index theorem with boundary (see for instance \cite{melroseAPSindex}). In this regard, a recent work of B\"ar and  Strohmaier proved a Lorentzian equivalent to Atiyah-Patodi-Singer for the Dirac equation \cite{bardirac}.
 
Even though closure is not achieved, the formula is expected to bring many secondary results. For instance, as discussed in the $2+1$ setting, we expect singularities and black holes to play a role in explaining the 
dark energy component in the standard model of cosmology. When applied to the complement of the horizon of a 
black hole, the boundary term in the formula happens to provide a radiative contribution to the black hole; however, this contribution turned out to be zero due to the symmetries of the spatially projected magnetic part of the Weyl tensor for Kerr-Newman black holes. It is still expected to be non-trivial for non-stationary black holes. 

Achievement of closure would have a far-reaching impact on various conundrums of contemporary cosmology. 
It would provide a model-independent approach to cosmological modeling from a general and global point of view. Many modeling assumptions for closure of the spatially averaged Einstein equations have been suggested (\textit{cf.} section~\ref{closureliterature}) that respect (i) the generic coupling of geometry (curvature) to the sources also for the averaged variables, while in FLRW cosmology the impact of inhomogeneities on the global model is neglected, (ii) the generic non-conservation of the average scalar curvature \cite{buchertcarfora:curvature}, while in FLRW cosmology curvature is assumed to obey a conservation law,  and (iii) the generic possibility of the change of sign of the averaged scalar curvature, being impossible in FLRW cosmology. There are show-cases that yield a natural and consistent explanation of (i) dark energy as a result of the coupling of structure formation to global properties of the universe model \cite{buchert:guadeloupe} and references therein, (ii) the coincidence problem, i.e. the fact that a dark energy component appears to become relevant for the universal expansion at the epoch of formation of nonlinear structures, (iii) a transition of positive initial curvature to a present-day negative curvature, (iv) the small matter density cosmological parameter found in local probes of the matter density, (v) the large angular diameter distance to the Cosmic Microwave Background consistent with supernova constraints, and (vi) the local expansion rate measurements (removal of the `Hubble tension'); for (i)-(vi) see the recent paper \cite{curvaturecrisis} and references therein. The closure assumption employed in \cite{curvaturecrisis} as well as others listed in \ref{closureliterature}, prescribe the average scalar curvature as a function of the volume. Such an assumption could then be verified without the need for sophisticated local modeling of structure formation or general-relativistic numerical simulations.
%
%
\ack
This work is part of a project that has received funding from the European Research Council (ERC) under the European Union's Horizon 2020 research and innovation programme (grant agreement ERC advanced grant 740021--ARTHUS, PI: TB). TB also acknowledges generous support and hospitality by the National Astronomical Observatory in Tokyo, as well as hospitality at Tohoku University in Sendai during 1999 and 2000, where the appendix material was written. We wish to thank Mauro Carfora for critically accompanying this work, George Ellis for his constant reminders of the need to extend the hierarchy of the averaged equations, Thierry Barbot, Martin France, Toshi Futamase, \'Eric Gourgoulhon, Asta Heinesen, \'Etienne Jaupart, Stefano Magni, Pierre Mourier, Jan Ostrowski, Pratyush Pranav, Syksy R\"as\"anen, Nezihe Uzun and Quentin Vigneron for valuable comments and discussions, Domenico Giulini, Pierre Mourier and Matthieu Chatelain for detailed comments on the manuscript. We thank the referees for their numerous constructive comments that helped us improve this paper.
We also wish to thank the \textit{SageMath} community, and especially \'Eric Gourgoulhon, for the development of \textit{SageManifolds} which has been used to check many computations. The software \textit{SageManifolds} \cite{sagemanifolds} is licensed as a free-software package under the GNU GPL, and is available at 
\url{https://sagemanifolds.obspm.fr/} .

\appendix

\section{On evolution equations for averaged curvature invariants and backreaction---more details on the closure problem}
\label{app:closure}

In this appendix we shall put unpublished results of an earlier draft\footnote{This appendix contains the essential material written in Sendai and Tokyo, Japan, in 1999-2000 as a follow-up of the invited paper \cite{buchert:jgrg}.} into perspective to discuss the issue of closure of the system of averaged equations, but also to explicitly demonstrate that the strategy to derive evolution equations for averaged curvature invariants and backreaction is not unique, i.e. the result largely depends on the strategy used. The purpose of these additional informations is to direct the reader to complementary insights while obtaining different evolutions equations as those obtained in the main text via the route of following the GBC approach. Since the problem of closure is not yet solved in this paper, but to our opinion more transparently formulated, the following may be also useful for further progress on this front. We emphasize that with these additional equations the closure problem is formulated at a deeper level of the hierarchy of evolution equations; imposing various closure conditions employed in the literature, \textit{cf.} section \ref{intro}, is thus possible on this deeper level, where
from a perturbative perspective we encounter several orders-of-magnitude smaller source terms, which points to a better control of approximations for the average dynamics on large scales.

In the following we shall refer to the set of $3+1$ Einstein equations for an irrotational dust matter model in a flow-orthogonal foliation of spacetime as in the main text, hence, all references to equations of the main text are applied to the case $N=3$. We write the Einstein equations in the following form (adopting the same conventions as in the main text, i.e. $8\pi G = c = 1$, except that we denote the first principal scalar invariant $\inI$ of the expansion tensor by the rate of expansion $\inI = \Theta$):
\begin{eqnarray}
\label{einsteindust}
\frac{1}{2} {\CR}
+ \frac{1}{3}\Theta^2 - \sigma^2  = \varrho + \Lambda \;\;\;;\;\;\;
{\sigma}^i_{\,\;j || i} = \frac{2}{3} \Theta_{| j} \;\;;\nonumber\\
\partial_t \varrho = - \Theta \varrho \;\;;\;\;
\partial_t {g}_{ij} = 2 \;\,{g}_{ik}{\sigma}^k_{\,\;j} 
+ \frac{2}{3}\Theta {g}_{ik}{\delta}^k_{\,\;j}\;\;;\nonumber \\
\partial_t \Theta + \frac{1}{3}\Theta^2 + 2\sigma^2 + \frac{\varrho}{2} - 
\Lambda \;=\;0 \;\;;\nonumber\\
\partial_t {\sigma}^i_{\,\;j} + \Theta {\sigma}^i_{\,\;j} = -
\left(\CRicR^{\,i}_{\;\,j} -\frac{1}{3}{\delta}^i_{\,\;j}{\CR}\right) =: - \tau^i_{\;\,j}\;\;,
\label{einstein}
\end{eqnarray}
where $\sigma_{ij}$ are the coefficients of shear, $\sigma^2 := 1/2 \sigma^i_{\;\,j} \sigma^j_{\;\,i}$ the rate of shear as in the main text, and $\tau_{ij}$ denote the trace-free part of the spatial Ricci tensor components $\CRicR_{\,ij}$.

We are going to employ the trace-parts of Eqs.~(\ref{einsteindust}), the energy constraint, the continuity equation, Raychaudhuri's equation 
and the last of Eqs.~(\ref{einsteindust}) from which we derive the following evolution equations for the 
rate of shear and the scalar curvature, respectively, by contracting this equation with the shear tensor. Also the last evolution equation can be easily verified:
\begin{eqnarray}
{\CR} +\frac{2}{3}\Theta^2 - 2\sigma^2 = 2\varrho + 2\Lambda \;\;;\nonumber \\
\dot\varrho + \Theta \varrho = 0 \;\;;\nonumber \\
\dot\Theta + \frac{1}{3}\Theta^2 + 2\sigma^2 = \Lambda - \frac{\varrho}{2}\;\;;\nonumber\\
2 (\sigma^2 \dot{)} + 4 \Theta \sigma^2 = \dot\CR + \frac{2}{3}\Theta \CR = - \nu \;\;;\;\;\nu : = 2 \tau^i_{\;\,j}\sigma^j_{\;\,i}\;\;;\nonumber\\
\dot\nu + \Theta \nu = \kappa - 4 \tau^2 \;\;;\;\;\kappa : = 2 \dot\tau^i_{\;\,j} \sigma^j_{\;\,i}\;\;.
\label{scalarparts}
\end{eqnarray}
An overdot abbreviates the partial time-derivative here, and we have introduced the rate of curvature anisotropies $\tau^2 := 1/2 \tau^i_{\;\,j} \tau^j_{\;\,i}$.
The above system of five equations (\ref{scalarparts}) for the seven scalar variables $\varrho$, $\Theta$, $\CR$, $\sigma$, $\tau$, $\nu$ and $\kappa$
is not closed; evolution equations for the fields $\tau$ and $\kappa$ are needed.

Going one level down the set of scalar evolution equations corresponds, in the averaged case, to finding evolution equations for the averaged scalar curvature $\average{\CR}$ and the kinematical backreaction $\CQ_\CD$.
Such a strategy turns out to be fairly involved, and not unique. The resulting form of the evolution equation will strongly depend on the way we derive them. In particular, we can `force' a particular form of the coefficients and subsequently obtain different source terms. On these grounds, the GBC approach in this paper provides a unique way of obtaining those evolution equations.

The first steps (presented in the next two subsections) are straightforward: we average the remaining evolution equations in the set (\ref{scalarparts}) that have not been explicitly used thus far.
We begin with the first set of equations which furnishes a useful addition to previous work.

\subsection{Re-deriving the integrability condition \eref{equ:cont}}

In previous work, the integrability condition \eref{equ:cont} has been shown to hold by comparing the derivative of the averaged energy constraint \eref{hamiltonav} with the averaged Raychaudhuri equation \eref{rayav}. The following shows that we can \textit{derive} this condition also from the set of $3+1$ Einstein equations. 
Consider the equations
\begin{eqnarray}
2 (\sigma^2\dot) + 4 \Theta \sigma^2 = -\nu \label{shearnu}   \;\;;\\
\dot{\CR} + \frac{2}{3}\Theta \CR = - \nu \;\;, \label{Rnu}
\end{eqnarray}
where $\nu : = 2 \tau^i_{\;\,j}\sigma^j_{\;\,i}$.
The corresponding averaged equations, written for the kinematical backreaction and averaged scalar curvature, read:
\begin{eqnarray}
{\dot\CQ}_\CD + 2\average{\Theta}\CQ_\CD - \average{\mu} = \average{\nu} \label{averageshearnu}\;\;;\\
\dotaverage{\CR} + \frac{2}{3}\average{\Theta}\average{\CR} + \average{\mu} = -\average{\nu}\;\;,\label{averageRnu}
\end{eqnarray}
where a new variable appeared as a result of non-commutativity of averaging and time-evolution: 
\begin{equation}
\average{\mu}: = -\frac{1}{3} \left( \average{\Theta\CR} - \average{\Theta}\average{\CR} \right) \;\;.
\end{equation}
The proof that Eq.~(\ref{averageshearnu}) follows from Eq.~(\ref{shearnu}) uses the definition of $\CQ_\CD$, {\it cf.} Eq.~\eref{QN}, here written as 
\begin{equation}
\CQ_\CD = \frac{2}{3} \left(\average{\Theta^2} - \average{\Theta}^2 \right) - 2 \average{\sigma}^2 \ ,
\end{equation}
and the local and averaged Raychaudhuri equations with the local and averaged energy constraints inserted
(together with a repeated use of the commutation rule \eref{eq:comrule}):\footnote{The fact that both
equations are equal, despite the non-commutativity of averaging and time-evolution, is non-trivial and due to the quadratic nonlinearity in $\Theta$ in Raychaudhuri's equation.}
\begin{eqnarray}
\dot{\Theta} + \Theta^2 = 3\Lambda + \frac{3}{2} \varrho - \CR \;\;;\label{theta}\\
\dotaverage{\Theta} + \average{\Theta}^2 = 3\Lambda + \frac{3}{2} \average{\varrho} - \average{\CR}\label{averagetheta} \;\;.
\end{eqnarray}
The new variable $\average{\mu}$ can be viewed to arise as the average of the following local quantities:
\begin{equation}
\mu_1 := -\frac{1}{3} \left(\delta\Theta \delta\CR \right)\;\;;\;\;\mu_2 := -\frac{1}{3}\left(\Theta \delta\CR\right)
\;\;;\;\;\mu_3 := -\frac{1}{3} \left(\CR\delta\Theta\right)  \;\;,
\end{equation}
where $\delta\Theta : = \Theta - \average{\Theta}$ and $\delta\CR := \CR - \average{\CR}$ denote the deviation fields of the local quantities with respect 
to their average. We can always use expressions of the type as the first of these expressions to provide an upper bound on the magnitude, here of $\average{\mu}$, using Schwarz' inequality:
\begin{equation}
|\average{\mu}| = |\average{\mu_1}|\;\;\le\;\;  \frac{1}{3} (\Delta \Theta \Delta \CR )^{1/2}\;\;, 
\end{equation}
with the variances (fluctuation amplitudes) $\Delta \Theta : = \average{(\delta\Theta)^2}$ and   $\Delta \CR : =\average{(\delta\CR)^2}$.
In the calculation below, only the last local expression $\mu_3$ will appear, and we therefore define $\mu$  throughout this appendix as follows:
\begin{equation}
\mu := \mu_3 = -\frac{1}{3} \left( \CR\delta\Theta \right)\;\;.
\end{equation}
We conclude that the equations (\ref{averageshearnu}, \ref{averageRnu}) reproduce the integrability condition \eref{equ:cont}. Moreover, it provides the further information on the nature of the sources for both sides of the integrability condition \eref{equ:cont}, i.e. with $\average{\Theta} = : 3 H_\CD$:
\begin{equation}
\label{explicitintegrability}
{\dot\CQ}_\CD + 6 H_\CD \CQ_\CD = - \left(\dotaverage{\CR}+ 2 H_\CD \average{\CR}\right) = \average{\mu} + \average{\nu} \;\;.
\end{equation}

\subsubsection*{Remark:}

The following is a trivial consequence of the derivation of Eq.~(\ref{explicitintegrability}), but it is useful as an illustration for the later investigations: 
the above additional information implies a {\it geometrical constraint} that has to hold on the $t-$hypersurfaces, \textit{if and only if} the averaged scalar curvature is restricted to evolve like a constant-curvature model, i.e. according to the solution $\average{\CR}^H = 6k / a_\D^2$ that, this latter, mirrors the situation in the standard FLRW cosmologies: 
\begin{equation}
\label{constantcurvatureconstraint1}
\average{\mu} + \average{\nu}=0\quad\Leftrightarrow\quad \frac{1}{3} \left( \average{\Theta\CR} - \average{\Theta}\average{\CR} \right) = 2 \average{\tau^i_{\;\,j}\sigma^j_{\;\,i}} \ .
\end{equation}
Rewriting the last term in terms of the full $3-$Ricci curvature $\CRicR_{\,ij} = 1/3 g_{ij} \CR + \tau_{ij}$ 
and the full expansion tensor $\Theta_{ij} = 1/3 g_{ij}\Theta + \sigma_{ij}$, we calculate 
\begin{equation}
\label{einsteinconstraint}
2 \average{\tau^i_{\;\,j}\sigma^j_{\;\,i}}= 2 \average{\CRicR^{\,i}_{\;\,j}\Theta^j_{\;\,i}} - \frac{2}{3} \average{\Theta\CR}
= -\langle{\dot\CR}\rangle_\CD - \frac{2}{3} \average{\Theta\CR}\;\;,
\end{equation}
where the last equality follows from Einstein's equations (\ref{einstein}). Thus,  we can use this general equation to show the inverse statement, namely that the above constraint 
(\ref{constantcurvatureconstraint1}) is equivalent to the evolution equation restricting our model to a constant-curvature model:    
\begin{equation}
\label{constantcurvatureconstraint2}
\fl
\average{\mu} + \average{\nu}=0\ \Leftrightarrow\  0=\langle{\dot\CR}\rangle_\CD + \average{\Theta\CR} -  \frac{1}{3} \average{\Theta}\average{\CR} =\dotaverage{\CR} + \frac{2}{3} \average{\Theta}\average{\CR},
\end{equation}
where in the last step we have again used the commutation rule \eref{eq:comrule}.

\subsection{Second-order evolution equations for $\CQ_\CD$ and $\average{\CR}$}

Employing also the last of the set of the equations (\ref{scalarparts}) carries us to second-order evolution equations for the variables
$\CQ_\CD$ and $\average{\CR}$. Formally, by taking the time-derivative of (\ref{explicitintegrability}), we obtain:
\begin{equation}
\fl
\ddot{\CQ}_\CD + 6 H_\CD {\dot\CQ}_\CD + 6 {\dot H}_\CD \CQ_\CD = -\left( \ddotaverage{\CR} + 2 H_\CD \dotaverage{\CR} + 2 {\dot H}_\CD
\average{\CR} \right) = \dotaverage{\nu} + \dotaverage{\mu} .
\end{equation}
The evolution equation for $\average{\nu}$ can be found straightforwardly, {\it cf.} (\ref{scalarparts}, \ref{eq:comrule}):
\begin{equation}
\average{{\dot\nu} + \Theta\nu} = \dotaverage{\nu} + 3 H_\CD \average{\nu} = \average{\kappa}-\average{4\tau^2} .
\end{equation}
Here, we do not worry about the appearance of the fields $\kappa$ and $\tau$, since we 
stop our considerations at this level of the hierarchy, i.e. we do not aim at including evolution equations for these fields, see (\ref{scalarparts}).
The calculation of the evolution of the average of the field $\mu$ only requires evolution equations that we have already used, i.e.
for the scalar curvature (\ref{Rnu}) and its average (\ref{averageRnu}), and for the rate of expansion (\ref{theta}) and its average (\ref{averagetheta}).
After a longer calculation we obtain:
\begin{eqnarray}  
\dotaverage{\mu} + 8 H_\CD \average{\mu} =
\frac{1}{3} \left( \average{\nu \Theta} - \average{\nu}\average{\Theta}\right) - \frac{2}{3} \left( \average{\mu \Theta} - \average{\mu}\average{\Theta}\right) \nonumber\\+ \frac{1}{3} \left( \average{\CR^2} - \average{\CR}^2
\right) - \frac{1}{2} \left( \average{\varrho \CR} - \average{\varrho}\average{\CR} \right)\;\;,
\end{eqnarray} 
where we have rewritten the term $\frac{2}{9}\left(\average{\Theta^2 \CR } - \average{\Theta}^2 \average{\CR}\right) = -\frac{4}{3}\average{\Theta}\average{\mu}
- \frac{2}{3} \left( \average{\mu \Theta} - \average{\mu}\average{\Theta}\right)$, using the local definition $\mu = -\frac{1}{3} \delta\Theta \CR$.

\medskip\noindent
We summarize the above general results in the following theorem.

\begin{theo}[Buchert]
\label{theoapp}

Under the same assumptions as in the main theorem \ref{theo:final}, the kinematical backreaction $\CQ_\CD$ and the average scalar curvature $\average{\CR}$ obey the following set of second-order evolution equations:
\begin{eqnarray}
\label{generalevolution}
\fl
\ddot{\CQ}_\CD + 6 H_\CD {\dot\CQ}_\CD + 6 {\dot H}_\CD \CQ_\CD = -\left( \ddotaverage{\CR} + 2 H_\CD \dotaverage{\CR} + 2 {\dot H}_\CD
\average{\CR} \right) = \dotaverage{\nu} + \dotaverage{\mu}\nonumber\\
\fl\qquad
= -3 H_D \average{\nu} + \frac{1}{3} \left( \average{\nu \Theta} - \average{\nu}\average{\Theta}\right) - 8 H_\CD  \average{\mu}
- \frac{2}{3} \left( \average{\mu \Theta} - \average{\mu}\average{\Theta}\right)\nonumber\\
\fl\qquad\quad + \CQ^\CR_\CD
- \frac{1}{2} \left( \average{\varrho \CR} - \average{\varrho}\average{\CR} \right)  +\average{\kappa}\;\;,
\end{eqnarray}
where we have defined the volume Hubble rate $H_\CD := \frac{1}{3} \average{\Theta}$, and the variables
\begin{eqnarray}
\label{intrinsicbackreaction}
\fl
\nu : = 2 \tau^i_{\;\,j} \sigma^j_{\;\,i}\;\;;\;\;\mu: = -\frac{1}{3} (\Theta - \average{\Theta}) \CR \;\;;\;\; \sigma^j_{\;\,i}: = \Theta^j_{\;\,i}- \frac{1}{3} \delta^j_{\;i}\Theta \;\;;\;\;\tau^i_{\;\,j}:= \CR^i_{\;\,j}
- \frac{1}{3} \delta^i_{\;j}\CR \;\;;
\nonumber\\
\fl
\CQ_\CD : =   \frac{2}{3} \left( \average{\Theta^2} - \average{\Theta}^2 \right) - 2\average{\sigma^2} \;\;;\;\;\sigma^2 := \frac{1}{2} \sigma^i_{\;\,j}\sigma^j_{\;\,i}\;\;;\nonumber\\ 
\fl
\CQ^\CR_\CD : =   \frac{1}{3} \left( \average{\CR^2} - \average{\CR}^2 \right) - 4\average{\tau^2 }
\;\;;\;\;\tau^2 := \frac{1}{2} \tau^i_{\;\,j}\tau^j_{\;\,i}\;\;;\;\;\kappa : = 2 \dot\tau^i_{\;\,j} \sigma^j_{\;\,i}\;\;,
\label{maineq}
\end{eqnarray}
and where the averaged dynamics of the volume Hubble rate $H_\CD$ couples to the variables  $\CQ_\CD$ and $ \average{\CR}$ according to the general cosmological equations:
\begin{equation}
\label{Hevolution}
3 H_\CD ^2 =  \varrho^{\CD}_{\rm eff} + \Lambda\;\;;\;\;
3 {\dot H}_\CD = -\frac{3}{2} (\varrho^{\CD}_{\rm eff}+{p}^{\CD}_{\rm eff}) \;\;,
\end{equation}
with the effective sources, 
\begin{equation}
\varrho^{\CD}_{\rm eff} =\average{\varrho} -\frac{1}{2}{\CQ}_\CD - \frac{1}{2}\average{\CR}\;\;;\;\;
{p}^{\CD}_{\rm eff} =  -\frac{1}{2}{\CQ}_\CD + \frac{1}{6}\average{\CR}\;,
\end{equation}
obeying the conservation law
${\dot\varrho}^{\CD}_{\rm eff} + 
3H_\CD   \left(\varrho^{\CD}_{\rm eff} +{p}^{\CD}_{\rm eff} \right)=0$ (\textit{cf.} section~\ref{averagedeqs}).
\end{theo}

\subsubsection*{Remarks:}

Note that the form of theorem \ref{theoapp} is such that the left-hand sides of Eq.~\eref{generalevolution} both vanish in the case of a FLRW model, as do the right-hand sides, since $\nu = \mu = \kappa= \sigma_{ij} = \tau_{ij} = 0$. 
The new backreaction term $\CQ^\CR_\CD$ appearing in Eq.~\eref{generalevolution}, and defined in Eq.~\eref{maineq}, has been named \textit{(intrinsic) curvature backreaction} and also arises in the context of Ricci flow smoothing of cosmological manifolds [Eq.~(116)]\cite{buchertcarfora}.


\section*{References}
\begin{thebibliography}{100}
\bibitem{alty}
Alty~L~J 1995
The generalized Gauss-Bonnet-Chern theorem
\emph{J. Math. Phys.} \href{http://dx.doi.org/10.1063/1.531015}{\textbf{36}, 3094}

\bibitem{Cosmologicaltime}
Andersson~L, Galloway~G~J and Howard~R 1998
The cosmological time function
\emph{Class. Quantum Grav.} \href{https://doi.org/10.1088/0264-9381/15/2/006}{\textbf{15} 309}
[\href{https://arxiv.org/abs/gr-qc/9709084}{arXiv:gr-qc/9709084}]

\bibitem{Avez}
Avez~A 1962, Formule de Gauss-Bonnet-Chern en m\'etrique de signature
quelconque  
\emph{C. R. Acad. Sci. Paris} \href{https://inmabb.criba.edu.ar/revuma/pdf/v21n4/p191-197.pdf}{\textbf{255} 2049}

\bibitem{bardirac}
B\"{a}r~C and Strohmaier~A 2019 
An index theorem for Lorentzian manifolds with compact spacelike Cauchy boundary
\emph{Amer. J. Math.} \href{https://doi.org/10.1353/ajm.2019.0037}{\textbf{141} 5}
[\href{https://arxiv.org/abs/1506.00959}{arXiv:1506.00959}]
  
\bibitem{1992PhRvL691849B}
Ba\~nados~M, Teitelboim~C and Zanelli~J 1992
Black hole in three-dimensional spacetime,
\emph{Phys. Rev. Lett.} \href{https://doi.org/10.1103/PhysRevLett.69.1849}{\textbf{69} 1849} 
[\href{https://arxiv.org/abs/hep-th/9204099}{arXiv:hep-th/9204099}]

\bibitem{zimdahl:viscosity}
Barbosa~R~M, Chirinos Isidro~E~G, Zimdahl~W and Piattella~O~F 2016
Cosmic bulk viscosity through backreaction
\emph{Gen. Relativ. Grav.} \href{https://doi.org/10.1007/s10714-016-2043-4}{\textbf{48} 51}
[\href{https://arxiv.org/abs/1512.07835}{arXiv:1512.07835}]

\bibitem{barbot_globally_2004}
Barbot~T 2005 
Globally hyperbolic flat spacetimes
\emph{J. Geom. Phys.} \href{https://doi.org/10.1016/j.geomphys.2004.05.002}{\textbf{53} 123}
[\href{https://arxiv.org/abs/math/0402257}{arXiv:math/0402257}]

\bibitem {Particules_1}
Barbot~T, Bonsante~F and Schlenker~J-M 2011 
Collisions of particles in locally AdS spacetimes I.  Local description and global examples
\emph{Comm. Math. Phys.} \href{https://doi.org/10.1007/s00220-011-1318-6}{\textbf{308} 147}
[\href{https://arxiv.org/abs/1010.3602}{arXiv:1010.3602}]

\bibitem{Particules_2}
Barbot~T, Bonsante~F and Schlenker~J-M 2014 
Collisions of particles in locally AdS spacetimes II. Moduli of globally hyperbolic spaces
\emph{Comm. Math. Phys.} \href{https://doi.org/10.1007/s00220-014-2020-2}{\textbf{327} 691}
[\href{https://arxiv.org/abs/1202.5753}{arXiv:1202.5753}]

\bibitem{MR3704814}
Belraouti~M 2017
Asymptotic behavior of Cauchy hypersurfaces in constant curvature space-times
\emph{Geom. Dedicata} \href{https://doi.org/10.1007/s10711-017-0230-4}{\textbf{190} 103}
[\href{https://arxiv.org/abs/1503.06343}{arXiv:1503.06343}]

\bibitem{MR2499272}
Benedetti~R and Bonsante~F 2009
Canonical Wick rotations in $3-$dimensional gravity 
\emph{Mem. Amer. Math. Soc.} \textit{AMS} \href{http://dx.doi.org/10.1090/memo/0926}{\textbf{198} 926}
[\href{https://arxiv.org/abs/math/0508485}{arXiv:math/0508485}]

\bibitem{latticesBH}
Bengtsson~I and Galstyan~I 2018
Black Hole Lattices Under the Microscope
\emph{Class. Quantum Grav.} \href{https://doi.org/10.1088/1361-6382/aac7e0}{\textbf{35} 145004}
[\href{https://arxiv.org/abs/1802.10396}{arXiv:1802.10396}]

\bibitem{latticesBH:review}
Bentivegna~E, Clifton~T, Durk~J, Korzy\'nski~M and Rosquist~K 2018 
Black-Hole Lattices as Cosmological Models. 
\emph{Class. Quantum Grav.} \href{https://doi.org/10.1088/1361-6382/aac846}{\textbf{35} 175004}
[\href{https://arxiv.org/abs/1801.01083}{arXiv:1801.01083}]

\bibitem{bergery}
Bergery~L~B 1981
La courbure scalaire des vari\'et\'es riemanniennes
\emph{S\'eminaire N. Bourbaki} \href{http://www.numdam.org/article/SB_1979-1980__22__225_0.pdf}{\textbf{556} 225}

\bibitem{MR867684}
Besse~A~L 1987 
Einstein manifolds
\emph{Ergebnisse der Mathematik und ihrer Grenzgebiete} \textbf{10},
\emph{Springer-Verlag, Berlin} \href{https://www.springer.com/de/book/9783540741206}{(ISBN 978-3-540-74311-8)}

\bibitem{bolejko}
Bolejko~K 2009
Volume averaging in the quasispherical Szekeres model
\emph{Gen. Relativ. Grav.} \href{https://doi.org/10.1007/s10714-008-0727-0}{\textbf{41} 1585}
[\href{https://arxiv.org/abs/0808.0376}{arXiv:0808.0376}]

\bibitem{bolejko:silent}
Bolejko~K 2017
Relativistic numerical cosmology with silent universes
\emph{Class. Quantum Grav.} \href{https://doi.org/10.1088/1361-6382/aa9d32}{\textbf{35} 024003} 
[\href{https://arxiv.org/abs/1708.09143}{arXiv:1708.09143}]

\bibitem{bolejko:review}
Bolejko~K, C\'el\'erier~ M-N and Krasi\'nski~A 2011
Inhomogeneous cosmological models: exact solutions and their applications 
\emph{Class. Quantum Grav.} \href{https://doi.org/10.1088/0264-9381/28/16/164002}{\textbf{28} 164002}
[\href{https://arxiv.org/abs/1102.1449}{arXiv:1102.1449}]

\bibitem{bonsante_flat_2003}
Bonsante~F 2005
Flat spacetimes with compact hyperbolic Cauchy surfaces
\emph{J. Diff. Geom.} \href{https://doi.org/10.4310/jdg/1122493997}{\textbf{69} 441}

\bibitem{MR3493421}
Bonsante~F and Seppi~A 2016
On Codazzi tensors on a hyperbolic surface and flat Lorentzian geometry
\emph{Int. Math. Res. Not.} \href{https://doi.org/10.1093/imrn/rnv144}{\textbf{2016} 343}
[\href{https://arxiv.org/abs/1501.04922}{arXiv:1501.04922}]

\bibitem{Boozer}
Boozer A D 2008
General relativity in ($1 + 1$) dimensions
\emph{European Journal of Physics} \href{http://dx.doi.org/10.1088/0143-0807/29/2/013}{\textbf{29} 2}

\bibitem{Bredon}
Bredon~G~E 1993
Topology and Geometry
\emph{Graduate Texts in Mathematics}, {\textit{Springer-Verlag, Berlin}}  \href{https://www.springer.com/de/book/9780387979267}{(ISBN 978-1-4757-6848-0)}

\bibitem{marco:silent}
Bruni~M, Matarrese~S and Pantano~O 1995
Dynamics of silent universes
\emph{Astrophys. J.} \href{http://dx.doi.org/10.1086/175755}{\textbf{445} 958}
[\href{https://arxiv.org/abs/astro-ph/9406068}{arXiv:astro-ph/9406068}]

\bibitem{brunswic:hal-01401821}
Brunswic~L 2016 
BTZ extensions of globally hyperbolic singular flat spacetimes
[\href{https://arxiv.org/abs/1605.05530}{arXiv:1605.05530}]

\bibitem{brunswic:hal-01317447}
Brunswic~L 2016 
Cauchy-compact flat spacetimes with BTZ singularities 
[\href{https://arxiv.org/abs/1611.08190}{arXiv:1611.08190}]

\bibitem{buchert:jgrg}
Buchert T 2000 
On average properties of inhomogeneous cosmologies,
In: 9th JGRG Meeting, Hiroshima 1999, Y. Eriguchi et al. (eds.), \emph{J.G.R.G.} \textbf{9}, 306 
[\href{https://arxiv.org/abs/gr-qc/0001056}{arXiv:gr-qc/0001056}]

\bibitem{buchert:dust}
Buchert~T 2000
On average properties of inhomogeneous fluids in general relativity I: Dust cosmologies
\emph{Gen. Relativ. Grav.} \href{http://dx.doi.org/10.1023/A:1001800617177}{\textbf{32} 105}
[\href{https://arxiv.org/abs/gr-qc/9906015}{arXiv:gr-qc/9906015}]

\bibitem{buchert:perfectfluid}
Buchert~T 2001
On average properties of inhomogeneous fluids in general relativity II: Perfect fluid cosmologies
\emph{Gen. Relativ. Grav.} \href{http://dx.doi.org/10.1023/A:1012061725841}{\textbf{33} 1381}
[\href{https://arxiv.org/abs/gr-qc/0102049}{arXiv:gr-qc/0102049}]

\bibitem{buchert:darkenergy}
Buchert~T 2005
A cosmic equation of state for the inhomogeneous universe: can a global far-from-equilibrium state explain dark energy?
\emph{Class. Quantum Grav.} \href{https://doi.org/10.1088/0264-9381/22/19/L01}{\textbf{22} L113}
[\href{https://arxiv.org/abs/gr-qc/0102049}{arXiv:gr-qc/0507028}]

\bibitem{buchert:global}
Buchert~T 2006
On globally static and stationary cosmologies with or without a cosmological constant and the dark energy problem
\emph{Class. Quantum Grav.} \href{https://doi.org/10.1088/0264-9381/23/3/017}{\textbf{23} 817}
[\href{http://arxiv.org/abs/gr-qc/0509124}{arXiv:gr-qc/0509124}]

\bibitem{buchert:review}
Buchert~T 2008
Dark Energy from structure: a status report
\emph{Gen. Relativ. Grav.} \href{https://doi.org/10.1007/s10714-007-0554-8}{\textbf{40} 467}
[\href{https://arxiv.org/abs/0707.2153}{arXiv:0707.2153}]

\bibitem{buchert:focus}
Buchert~T 2011
Toward physical cosmology: focus on inhomogeneous geometry and its non-perturbative effects
\emph{Class. Quantum Grav.} \href{https://doi.org/10.1088/0264-9381/28/16/164007}{\textbf{28} 164007}
[\href{https://arxiv.org/abs/1103.2016}{arXiv:1103.2016}]

\bibitem{buchert:guadeloupe}
Buchert~T 2019
Is Dark Energy Simulated by Structure Formation in the Universe?
In \emph{2nd World Summit on Exploring the Dark Side of the Universe, 25-29 June 2018, University of Antilles, Guadeloupe} \href{https://pos.sissa.it/cgi-bin/reader/conf.cgi?confid=335}{PoS(EDSU2018)\textbf{038}}
[\href{https://arxiv.org/abs/1810.09188}{arXiv:1810.09188}]

\bibitem{buchertcarfora}
Buchert~T and Carfora~M 2002
Regional averaging and scaling in relativistic cosmology
\emph{Class. Quantum Grav.} \href{https://doi.org/10.1088/0264-9381/19/23/314}{\textbf{19} 6109}
[\href{https://arxiv.org/abs/gr-qc/0210037}{arXiv:gr-qc/0210037}]

\bibitem{buchertcarfora:curvature}
Buchert~T and Carfora~M 2008
On the curvature of the present-day Universe
\emph{Class. Quantum Grav.} \href{https://doi.org/10.1088/0264-9381/25/19/195001}{\textbf{25} 195001}
[\href{https://arxiv.org/abs/0803.1401}{arXiv:0803.1401}]

\bibitem{buchert11}
Buchert~T, Carfora~M, Ellis~G~F~R, Kolb~E~W, MacCallum~M~A~H, Ostrowski~J~J, R\"as\"anen~S, Roukema~B~F, Andersson~L, Coley~A~A and Wiltshire~D~L 2015
Is there proof that backreaction of inhomogeneities is irrelevant in cosmology?
\emph{Class. Quantum Grav.} \href{https://doi.org/10.1088/0264-9381/32/21/215021}{\textbf{32}, 215021}
[\href{https://arxiv.org/abs/1505.07800}{arXiv:1505.07800}]  

\bibitem{tensions}
Buchert~T, Coley~A, Kleinert~H, Roukema~B~F and Wiltshire~D~L 2016
Observational challenges for the standard FLRW model
\emph{Int. J. of Mod. Phys. D} \href{http://dx.doi.org/10.1142/S021827181630007X}{\textbf{25} 1630007}
[\href{https://arxiv.org/abs/1512.03313}{arXiv:1512.03313}]

\bibitem{buchertehlers}
Buchert~T and Ehlers~J 1997
Averaging inhomogeneous Newtonian cosmologies
\emph{Astron. Astrophys.} \href{http://adsabs.harvard.edu/abs/1997A\%26A...320....1B}{\textbf{320} 1}
[\href{https://arxiv.org/abs/astro-ph/9510056}{arXiv:astro-ph/9510056}]

\bibitem{morphon}
Buchert~T, Larena~J and Alimi~J-M 2006
Correspondence between kinematical backreaction and scalar field cosmologies---the `morphon field'
\emph{Class. Quantum Grav.} \href{https://doi.org/10.1088/0264-9381/23/22/018}{\textbf{23} 6379}
[\href{https://arxiv.org/abs/gr-qc/0606020}{arXiv:gr-qc/0606020}]

\bibitem{foliations}
Buchert~T, Mourier~P and Roy~X 2018
On cosmological backreaction and its dependence on spacetime foliation
\emph{Class. Quantum Grav.} \href{https://doi.org/10.1088/1361-6382/aaebce}{\textbf{35} 24LT02}
[\href{https://arxiv.org/abs/1805.10455}{arXiv:1805.10455}]

\bibitem{buchert:generalfluid}
Buchert~T, Mourier~P and Roy~X 2020
On average properties of inhomogeneous fluids in general relativity III: general fluid cosmologies
\emph{Gen. Relativ. Grav.} \href{https://doi.org/10.1007/s10714-020-02670-6}{\textbf{52} 27}
[\href{https://arxiv.org/abs/1912.04213}{arXiv:1912.04213}]

\bibitem{lagpenrose}
Buchert~T and Mourier~P 2020
Lagrangian theory of structure formation in relativistic cosmology. VII. Information entropies, backreaction, and the Penrose conjecture, \emph{in preparation.} 

\bibitem{RZA_1}
Buchert~T and Ostermann~M 2012
Lagrangian theory of structure formation in relativistic cosmology. I. Lagrangian framework and definition of a nonperturbative approximation
\emph{Phys. Rev. D} \href{https://doi.org/10.1103/PhysRevD.86.023520}{\textbf{86} 023520}
[\href{http://arxiv.org/abs/arXiv:1203.6263}{arXiv:1203.6263}]

\bibitem{RZA_2}
Buchert~T, Nayet~C and Wiegand~A 2013
Lagrangian theory of structure formation in relativistic cosmology. II. Average properties of a generic evolution model
\emph{Phys. Rev. D} \href{https://doi.org/10.1103/PhysRevD.87.123503}{\textbf{87} 123503}
[\href{http://arxiv.org/abs/arXiv:1303.6193}{arXiv:1303.6193}]

\bibitem{inflation}
Buchert~T and Obadia~N 2011
Effective inhomogeneous inflation: curvature inhomogeneities of the Einstein vacuum
\emph{Class. Quantum Grav.} \href{https://doi.org/10.1088/0264-9381/28/16/162002}{\textbf{28} 162002}
[\href{http://arxiv.org/abs/arXiv:1010.4512}{arXiv:1010.4512}]

\bibitem{buchertrasanen}
Buchert~T and R{\"a}s{\"a}nen~S 2012
Backreaction in Late-Time Cosmology
\emph{Annu. Rev. Nucl. Part. Sci.} \href{https://doi.org/10.1146/annurev.nucl.012809.104435}{\textbf{62} 57}
[\href{http://arxiv.org/abs/arXiv:1112.5335}{arXiv:1112.5335}]

\bibitem{carlip}
Carlip~S 1998
\href{https://doi.org/10.1017/CBO9780511564192} {Quantum gravity in $2+1$ dimensions}
\textit{Cambridge Monographs on Math. Phys.}, Cambridge University Press

\bibitem{Chern1}
Chern~S 1944
A simple intrinsic proof of the Gauss-Bonnet formula for closed Riemannian manifolds
\emph{Ann. Math.} \href{https://doi.org/10.2307/1969302}{\textbf{45} 747}

\bibitem{MR155261}
Chern~S 1963
Pseudo-Riemannian geometry and the Gauss-Bonnet formula
\emph{An. Acad. Brasil. Ci.} \href{https://duetosymmetry.com/files/An.%20Acad.%20Brasil.%20Ci.%201963%20Chern.pdf}{\textbf{35} 31}

\bibitem{zimdahl:LTB}
Chirinos Isidro~E~G, Barbosa~R~M, Piattella~O~F and Zimdahl~W 2016
Averaged Lema\^\i tre-Tolman-Bondi dynamics
\emph{Class. Quantum Grav.} \href{https://doi.org/10.1088/1361-6382/34/3/035001}{\textbf{34} 035001}
[\href{https://arxiv.org/abs/1608.00452}{arXiv:1608.00452}]

\bibitem{Christodoulou}
Christodoulou~D 2009
\emph{EMS Monographs} \href{https://doi.org/10.4171/068}{The formation of black holes in general relativity} European Mathematical Society, Publishing House Switzerland

\bibitem{ellisreview}
Clarkson~C~A, Ellis~G~F~R, Larena~J and Umeh~O~C  2011
Does the growth of structure affect our dynamical models of the universe? The averaging, backreaction and fitting problems in cosmology
\emph{Rep. Prog. Phys.} \href{https://doi.org/10.1088/0034-4885/74/11/112901}{\textbf{74} 112901}
[\href{https://arxiv.org/abs/1109.2314}{arXiv:1109.2314}]

\bibitem{celiaSN}
Desgrange~C, Heinesen~A and Buchert~T 2019
Dynamical spatial curvature as a fit to type Ia supernovae
\emph{Int. J. of Mod. Phys. D} \href{https://doi.org/10.1142/S0218271819501438}{\textbf{28} 1950143}
[\href{https://arxiv.org/abs/1902.07915}{arXiv:1902.07915}]

\bibitem{ellisbuchert}
Ellis~G~F~R  and Buchert~T 2005
The universe seen at different scales
\emph{Phys. Lett. A} \href{https://doi.org/10.1016/j.physleta.2005.06.087}{\textbf{347} 38}
[\href{https://arxiv.org/abs/gr-qc/0506106}{arXiv:gr-qc/0506106}]

\bibitem{henk:silent}
van Elst~H, Uggla~C, Lesame~W~M, Ellis~G~F~R and Maartens~R 1997
Integrability of irrotational silent cosmological models
\emph{Class. Quantum Grav.} \href{https://doi.org/10.1088/0264-9381/14/5/018}{\textbf{14} 1151}
[\href{https://arxiv.org/abs/gr-qc/9611002}{arXiv:gr-qc/9611002}]

\bibitem{fennen2019lie}
Fennen~M and Giulini~D 2020
Lie sphere geometry in lattice cosmology
\emph{Class. Quantum Grav.} \href{https://doi.org/10.1088/1361-6382/ab6a20}{\textbf{37} 065007} 
[\href{https://arxiv.org/abs/1909.08109}{arXiv:1909.08109}]

\bibitem{galloway}
Galloway~G~J 1993
On the topology of black holes
\emph{Commun. Math. Phys.} \href{https://projecteuclid.org/euclid.cmp/1104252045}{\textbf{151}, 53--66}

\bibitem{Geroch}
Geroch~R 1970
Domain of Dependence
\emph{J. of Math. Phys.} \href{https://doi.org/10.1063/1.1665157}{\textbf{11} 2}

\bibitem{sagemanifolds}
Gourgoulhon~E, Bejger~M and Mancini~M 2015
Tensor calculus with open-source software: the \textit{SageManifolds} project 
\emph{J. Phys: Conf. Ser.} \href{https://doi.org/10.1088/1742-6596/600/1/012002}{\textbf{600} 012002}
[\href{https://arxiv.org/abs/1412.4765}{arXiv:1412.4765}]

\bibitem{Gregoris}
Gregoris~D, Ong~Y~C and Wang~B 2019
Curvature invariants and lower dimensional black hole horizons 
\emph{Eur. Phys. J. C} \href{https://doi.org/10.1140/epjc/s10052-019-7423-y}{\textbf{79} 925}
[\href{https://arxiv.org/abs/1902.05565}{arXiv:1902.05565}]

\bibitem{pak}
Grigorchuk~R and Pak~I 2006
Groups of intermediate growth: an introduction for beginners
[\href{https://arxiv.org/abs/math/0607384}{arXiv:math/0607384}]

\bibitem{hawking}
Hawking~S~W 1971
Black holes in general relativity
\emph{Commun. Math. Phys.} \href{https://projecteuclid.org/euclid.cmp/1103857884}{\textbf{25}, 152--166}

\bibitem{covariance}
Heinesen~A, Mourier~P and Buchert~T 2019
On the covariance of scalar averaging and backreaction in relativistic inhomogeneous cosmology
\emph{Class. Quantum Grav.} \href{https://doi.org/10.1088/1361-6382/ab0618}{\textbf{36} 075001}
[\href{https://arxiv.org/abs/1811.01374}{arXiv:1811.01374}]

\bibitem{curvaturecrisis}
Heinesen~A and Buchert~T 2020
Solving the curvature and Hubble parameter inconsistencies through structure formation-induced curvature
\emph{Class. Quantum Grav.} \href{https://doi.org/10.1088/1361-6382/ab954b}{\textbf{37} 164001} (Focus issue on the Hubble constant tension)
[\href{https://arxiv.org/abs/2002.10831}{arXiv:2002.10831}]

\bibitem{Huber}
Huber~A 1957
On subharmonic functions and differential geometry in the large 
\emph{Comment. Math. Helv.} \href{https://doi.org/10.1007/BF02564570}{\textbf{32} 13--72}

\bibitem{HT}
Hughston~L~P and Tod~K~P 1991
\href{https://doi.org/10.1017/CBO9781139171977}{An Introduction to General Relativity}, Cambridge Univ. Press

\bibitem{Troyanov}
Hulin~D, Troyanov~M 1992
Prescribing curvature on open surfaces 
\emph{Math. Ann.} \href{https://doi.org/10.1007/BF01444716}{\textbf{293} 2}

\bibitem{jackiwlowgravity}
Jackiw R 1985 Lower dimensional gravity
\emph{Nucl. Phys. B} \href{https://doi.org/10.1016/0550-3213(85)90448-1}{\textbf{252} 343}

\bibitem{kazdan}
Kazdan~J~L Warner~F~W 1975
\emph{Inventiones math.} \href{https://doi.org/10.1007/BF01425558}{\textbf{28} 227}

\bibitem{Kobayashi-Nomizu1}
Kobayashi~S and Nomizu~K 1996
\href{https://www.wiley.com/en-us/Foundations+of+Differential+Geometry%2C+Volume+1-p-9780471157335}{Foundations of Differential Geometry Volume 1} John Wiley \& Sons, New York

\bibitem{krazinski}
Krasi\'nski~A. 1997 
\href{https://www.cambridge.org/us/catalogue/catalogue.asp?isbn=0521481805}{Inhomogeneous Cosmological Models} \emph{Cambridge University Press} U.K.

\bibitem{Eric:BH}
Lamy~F, Gourgoulhon~E., Paumard~T and Vincent~F~H 2018
Imaging a non-singular rotating black hole at the center of the Galaxy 
\emph{Class. Quantum Grav.} \href{https://doi.org/10.1088/1361-6382/aabd97}{\textbf{35} 115009}
[\href{https://arxiv.org/abs/1802.01635}{arXiv:1802.01635}]

\bibitem{lanczos}
Lanczos~C 1938
A remarkable property of the Riemann-Christoffel tensor in four dimensions
\emph{Annals of Mathematics} \href{https://www.jstor.org/stable/1968467}{\textbf{39} 842}

\bibitem{DEtopology1}
Le Delliou~M and Lorca Espiro~J 2020 
Is the cosmological constant of topological origin?
\emph{Physics of the Dark Universe} \href{https://doi.org/10.1016/j.dark.2020.100569}{\textbf{29} 100569}
[\href{https://arxiv.org/abs/1906.03041}{arXiv:1906.03041}]

\bibitem{Weylentropy}
Li~N, Buchert~T, Hosoya~A, Morita~M and Schwarz~D~J 2012
Relative information entropy and Weyl curvature of the inhomogeneous Universe
\emph{Phys. Rev. D} \href{https://doi.org/10.1103/PhysRevD.86.083539}{\textbf{86} 083539}
[\href{https://arxiv.org/abs/1208.3376}{arXiv:1208.3376}]

\bibitem{lohkamp}
Lohkamp~J 1994
Metrics of negative Ricci curvature
\emph{Annals of Mathematics} \href{https://www.jstor.org/stable/2118620}{\textbf{140} 3}

\bibitem{DEtopology2}
Lorca Espiro~J and Le Delliou~M 2020
Dark energy from topology
\emph{J. Cosmol. Astropart. Phys.} \href{https://doi.org/10.1088/1475-7516/2020/03/020}{\textbf{03} 020} 
[\href{https://arxiv.org/abs/1910.03639}{arXiv:1910.03639}]

\bibitem{Magni}
Magni~S 2012
Backreaction and the covariant formalism of general relativity
\emph{Master Thesis (supervised by Syksy R\"as\"anen and Mauro Carfora)}, University of Pavia, Italy
[\href{https://arxiv.org/abs/1202.0430}{arXiv:1202.0430}]

\bibitem{melroseAPSindex}
Melrose~R~B 1993 
\href{https://www.crcpress.com/The-Atiyah-Patodi-Singer-Index-Theorem/Melrose/p/book/9781568810027}
{The Atiyah-Patodi-Singer index theorem}
\emph{Research notes in mathematics}
A K Peters/CRC Press

\bibitem{Mess}
Mess~G 2007 
Lorentz spacetimes of constant curvature
\emph{Geom. Dedicata} \href{https://doi.org/10.1007/s10711-007-9155-7}{\textbf{126} 3}
[\href{https://arxiv.org/abs/0706.1570}{arXiv:0706.1570}]

\bibitem{maeda:quiet}
Mutoh~H, Hirai~T and Maeda~K 1997
Dynamics of quiet universes
\emph{Phys. Rev. D} \href{https://doi.org/10.1103/PhysRevD.55.3276}{\textbf{55} 3276}
[\href{https://arxiv.org/abs/astro-ph/9608183}{arXiv:astro-ph/9608183}]

\bibitem{Nicolaescu}
Nicolaescu~L~I 2003
\href{https://doi.org/10.1515/9783110198102}{The Reidemeister Torsion of 3--Manifolds}
\textit{De Gruyter Studies in Mathematics} \textbf{30}, De Gruyter

\bibitem{jantopacc}
Ostrowski~J~J, Roukema~B~F and Buli\'nski~Z~P 2012
A relativistic model of the topological acceleration effect 
\emph{Class. Quantum Grav.} \href{https://doi.org/10.1088/0264-9381/29/16/165006}{\textbf{29} 165006}
[\href{https://arxiv.org/abs/1109.1596}{arXiv:1109.1596}]
 
\bibitem{PKbook}
Pleba\'nski~J and Krasi\'nski~A 2010
\href{https://doi.org/10.1017/CBO9780511617676}{An Introduction to General Relativity and Cosmology}
Cambridge University Press

\bibitem{MR143186}
Richards~I 1963
On the classification of noncompact surfaces
\emph{Trans. Amer. Math. Soc.} \href{https://doi.org/10.2307/1993768}{\textbf{106} 259}

\bibitem{romero}
Romero~C and Dahia~F 1994 
Theories of gravity in $2+1$ dimensions
\emph{Int. J. Theor. Phys.} \href{https://doi.org/10.1007/BF00675174}{\textbf{33} 2091}

\bibitem{boudtopacc1}
Roukema~B~F, Bajtlik~S, Biesiada~M, Szaniewska~A and Jurkiewicz~H 2007
A weak acceleration effect due to residual gravity in a multiply connected universe
\emph{Astron. Astrophys.} \href{https://doi.org/10.1051/0004-6361:20064979}{\textbf{463} 861}
[\href{https://arxiv.org/abs/astro-ph/0602159}{arXiv:astro-ph/0602159}]

\bibitem{boudtopacc2}
Roukema~B~F and R\'o$\dot z$a\'nski~P~T 2009
The residual gravity acceleration effect in the Poincar\'e dodecahedral space 
\emph{Astron. Astrophys.} \href{https://doi.org/10.1051/0004-6361/200911881}{\textbf{502} 27}
[\href{https://arxiv.org/abs/0902.3402}{arXiv:0902.3402}]

\bibitem{boudLTB}
Roukema~B~F, Blanl{\oe}il~V and Ostrowski~J~J 2013
Topological implications of inhomogeneity 
\emph{Phys. Rev. D} \href{https://doi.org/10.1103/PhysRevD.87.043521}{\textbf{87} 043521}
[\href{http://arxiv.org/abs/1201.5845}{arXiv:1201.5845}]

\bibitem{boudjan}
Roukema~B~F and Ostrowski~J~J 2019
Does spatial flatness forbid the turnaround epoch of collapsing structures? 
\emph{J. Cosmol. Astropart. Phys.} \href{https://doi.org/10.1088/1475-7516/2019/12/049}{\textbf{12} 049} 
[\href{http://arxiv.org/abs/1902.09064}{arXiv:1902.09064}]

\bibitem{chaplygin}
Roy~X and Buchert~T 2010
Chaplygin gas and effective description of inhomogeneous universe models in general relativity
\emph{Class. Quantum Grav.} \href{https://doi.org/10.1088/0264-9381/27/17/175013}{\textbf{27} 175013}
[\href{https://arxiv.org/abs/0909.4155}{arXiv:0909.4155}]

\bibitem{scaling}
Roy~X, Buchert~T, Carloni~S and Obadia~N 2011
Global gravitational instability of FLRW backgrounds---interpreting the dark sectors
\emph{Class. Quantum Grav.} \href{https://doi.org/10.1088/0264-9381/28/16/165004}{\textbf{28} 165004}
[\href{https://arxiv.org/abs/1103.1146}{arXiv:1103.1146}]

\bibitem{frankBH}
Steiner~F 2016
Do black holes exist in a finite Universe having the topology of a flat 3-torus? 
In: \emph{Ulmer Seminare 2016/2017, Funktionalanalysis und Evolutionsgleichungen} Heft \textbf{20}, 331-351 (Editors W. Arendt et al.), Institute of Applied Analysis, Ulm University, Ulm Germany 
[\href{https://arxiv.org/abs/1608.03133}{arXiv:1608.03133}]

\bibitem{sussman1}
Sussman~R~A 2011
Back-reaction and effective acceleration in generic LTB dust models
\emph{Class. Quantum Grav.} \href{https://doi.org/10.1088/0264-9381/28/23/235002}{\textbf{28} 235002}
[\href{https://arxiv.org/abs/1102.2663}{arXiv:1102.2663}]

\bibitem{sussman2}
Sussman~R~A 2013
Weighed scalar averaging in LTB dust models, part I: statistical fluctuations and gravitational entropy
\emph{Class. Quantum Grav.} \href{https://doi.org/10.1088/0264-9381/30/6/065015}{\textbf{30} 065015}
[\href{https://arxiv.org/abs/1209.1962}{arXiv:1209.1962}]

\bibitem{sussman3}
Sussman~R~A, Hidalgo~J~C, Delgado Gaspar~I and Germ\`an~G 2018
Nonspherical Szekeres models in the language of cosmological perturbations
\emph{Phys. Rev. D} \href{https://doi.org/10.1103/PhysRevD.95.064033}{\textbf{95} 064033}
[\href{https://arxiv.org/abs/1701.00819}{arXiv:1701.00819}]

\bibitem{szekeres1975class}
Szekeres~P 1975
A class of inhomogeneous cosmological models
\emph{Comm. Math. Phys.} \href{https://doi.org/10.1007/BF01608547}{\textbf{41} 55}

\bibitem{quentin:darkmatter}
Vigneron~Q and Buchert~T 2019
Dark matter from backreaction? Collapse models on galaxy cluster scales
\emph{Class. Quantum Grav.} \href{https://doi.org/10.1088/1361-6382/ab32d1}{\textbf{36} 175006} 
[\href{https://arxiv.org/abs/1902.08441}{arXiv:1902.08441}]

\bibitem{whatisdust}
Wiltshire~D~L 2011
What is dust?---Physical foundations of the averaging problem in cosmology
\emph{Class. Quantum Grav.} \href{https://doi.org/10.1088/0264-9381/28/16/164006 }{\textbf{28} 164006}
[\href{https://arxiv.org/abs/1106.1693}{arXiv:1106.1693}]

\bibitem{yodzis}
Yodzis~P 1974
On the expansion of closed universes
\emph{Proc. Royal Irish Acad.} \href{https://www.jstor.org/stable/20488732?seq=1/subjects}{\textbf{74A} 61}

\end{thebibliography}
\end{document}